\documentclass[journal]{IEEEtran}
\usepackage{soul}
\usepackage{algorithm}
\usepackage[noend]{algpseudocode}
\usepackage{enumitem}
\usepackage{graphicx}
\usepackage{mathtools}
\usepackage{amsmath}
\usepackage{amsfonts}
\usepackage{array}
\newcommand{\PreserveBackslash}[1]{\let\temp=\\#1\let\\=\temp}
\newcolumntype{C}[1]{>{\PreserveBackslash\centering}p{#1}}
\newcolumntype{R}[1]{>{\PreserveBackslash\raggedleft}p{#1}}
\newcolumntype{L}[1]{>{\PreserveBackslash\raggedright}p{#1}}
\usepackage[usenames]{color}
\usepackage{colortbl,booktabs}
\usepackage{tabularx}
\usepackage{multicol}
\usepackage{booktabs}
\usepackage[Symbol]{upgreek}
\usepackage{subfigure}
\usepackage{bm}
\usepackage{url}
\usepackage[usenames,dvipsnames,svgnames,table]{xcolor}

\usepackage{cite}
\usepackage{balance}
\usepackage{times}
\newtheorem{lemma}{Lemma}

\usepackage{stfloats}
\usepackage{balance}

\usepackage[nonumberlist,acronym,shortcuts]{glossaries}
\makeglossaries
\makeindex
\makeatletter
\g@addto@macro\normalsize{%
  \setlength\abovedisplayskip{5pt}
  \setlength\belowdisplayskip{5pt}
  \setlength\abovedisplayshortskip{5pt}
  \setlength\belowdisplayshortskip{5pt}
}
\makeatother

\begin{document}

\title{\huge Boosting Fronthaul Capacity: Global Optimization of Power Sharing
 for Centralized Radio Access Network}
\author{Jiankang~Zhang,~\IEEEmembership{Senior~Member,~IEEE},
        Sheng~Chen,~\IEEEmembership{Fellow,~IEEE},
        Xinying~Guo,
	Jia~Shi,
        Lajos~Hanzo,~\IEEEmembership{Fellow,~IEEE}
\thanks{J.~Zhang, S.~Chen and L.~Hanzo are with School of Electronics and Computer
 Science, University of Southampton, U.K. (E-mails: jz09v@ecs.soton.ac.uk, sqc@ecs.soton.ac.uk,
 lh@ecs.soton.ac.uk). S. Chen is also with King Abdulaziz University, Jeddah, Saudi Arabia.} %
\thanks{X.~Guo is with College of Information Science and Engineering, Henan University
 of Technology, China (E-mail: guoxinying@haut.edu.cn).} %
\thanks{J.~Shi is with the Integrated Service Networks Lab, Xidian University, Xi'an,
 China (E-mail: jiashi@xidian.edu.cn).} %
\thanks{The financial support of the EPSRC projects EP/Noo4558/1, EP/PO34284/1, of the Royal Society's GRFC Grant, and of the European Research Council's Advanced Fellow Grant QuantCom, as well as in part by the National
Natural Science Foundation of China under Grant 61571401.} %
\vspace*{-8mm}
}

\maketitle

\IEEEpeerreviewmaketitle

\begin{abstract}
 The limited fronthaul capacity imposes a challenge on the uplink of centralized
 radio access network (C-RAN). We propose to boost the fronthaul capacity of
 massive multiple-input multiple-output (MIMO) aided C-RAN by globally optimizing
 the power sharing between channel estimation and data transmission both for the
 user devices (UDs) and the remote radio units (RRUs). Intuitively, allocating
 more power to the channel estimation will result in more accurate channel estimates,
 which increases the achievable throughput. However, increasing the power allocated
 to the pilot training will reduce the power assigned to data transmission, which
 reduces the achievable throughput. In order to optimize the powers allocated to
 the pilot training and to the data transmission of both the UDs and the RRUs, we
 assign an individual power sharing factor to each of them and derive an asymptotic
 closed-form expression of the signal-to-interference-plus-noise for the massive
 MIMO aided C-RAN consisting of both the UD-to-RRU links and the RRU-to-baseband
 unit (BBU) links. We then exploit the C-RAN architecture's central computing and
 control capability for jointly optimizing the UDs' power sharing factors and the
 RRUs' power sharing factors aiming for maximizing the fronthaul capacity. Our
 simulation results show that the fronthaul capacity is significantly boosted by the
 proposed global optimization of the power allocation between channel estimation and
 data transmission both for the UDs and for their host RRUs.  As a specific example
 of 32 receive antennas (RAs) deployed by RRU and 128 RAs deployed by BBU, the
 sum-rate of 10 UDs achieved with the optimal power sharing factors improves 33\%
 compared with the one attained  without optimizing power sharing factors.
\end{abstract}

\begin{IEEEkeywords}
 Fronthaul, small cell, massive MIMO, C-RAN, power allocation.
\end{IEEEkeywords}

\section{Introduction}\label{S1}

 The fifth-generation (5G) wireless communication system is expected to meet the
 ever-increasing mobile traffic, predicted to be 291.8 Exabytes by 2019
 \cite{rappaport2015wideband}. The number of mobile user devices (UDs) of 5G
 connections is forecast to reach a figure between 25 million and 100 million
 by 2021 \cite{statista20185g}. In order to provide ubiquitous service access for
 such huge number of UDs, the small cell concept \cite{jungnickel2014therole}
 combined with massive multiple-input multiple-output (MIMO)
 \cite{marzetta2010noncooperative} and/or millimeter-wave (mmWave) technologies
 \cite{xiao2017millimeter} has emerged as one of the most promising network
 structures for 5G systems. However, substantial processing power is required for
 jointly exploiting the small cell's spatial diversity and temporal diversity.
 This motivates the concept of centralized radio access network (C-RAN)
 \cite{rost2014cloud,bartelt2015fronthaul}, which consists of baseband units (BBUs)
 and remote radio units (RRUs). In a C-RAN, multiple RRUs are connected to a BBU,
 which carries out all the baseband signal processing centrally, whilst the RRUs
 handle the radio frequency processing. Consequently, substantial amount of
 information is  exchanged over fronthaul links between RRUs and their host BBU,
 which imposes a bottleneck on C-RAN \cite{peng2017cost} and prevents its
 large-scale practical deployment.  

\subsection{Related Works}\label{S1.1}

 Extensive efforts have been devoted to the enhancement of fronthaul capacity,
 including compression/quantization
 \cite{zhou2016fronthaul,lee2016multivariate,vu2017adaptive}, quality of service
 (QoS) guarantee \cite{zhao2016cluster,lee2018dynamic}, interference mitigation
 \cite{liu2017compressive,hao2018price-based}, user/access link selection
 \cite{pan2017joint,ren2017joint,luong2017optimal}, as well as resource allocation
 and optimization \cite{peng2017cost,liu2015joint,peng2015fronthaul,hao2018small}. 

 To elaborate, compression/quantization schemes were designed for reducing the
 data traffic of fronthaul links to meet their capacity constraint. Explicitly,
 Zhou \emph{et al.} \cite{zhou2016fronthaul} studied a joint fronthaul compression
 and transmit beamforming scheme relying on noise covariance matrix quantization
 in the context of traditional MIMO having a small number of antennas. Lee
 \emph{et al.} \cite{lee2016multivariate} investigated multivariate fronthaul
 quantization motivated by the network-information theory, which is capable of
 jointly optimizing the downlink precoding and quantization at a reduced-complexity.
 By contrast, Vu \emph{et al.} \cite{vu2017adaptive} derived a block error rate
 metric in the context of Rayleigh fading channels for designing an adaptive
 compression scheme.

 QoS represents a general terminology that includes packet loss ratio, bit error
 ratio (BER), throughput, transmission delay, etc. Since it is challenging to
 investigate all the QoS metrics jointly, the existing studies mainly concentrate
 on just a single or a few aspects of QoS. Explicitly, by relying on content
 caching, Zhao \emph{et al.} \cite{zhao2016cluster} improved the link-level
 effective capacity, while the dynamic network slicing scheme of \cite{lee2018dynamic}
 improved the QoS in terms of rate-fairness, as well as maximum and minimum rates.

 Through interference reduction, the achievable fronthaul capacity can also be
 increased. Hence Liu \emph{et al.} \cite{liu2017compressive} focused the
 attention on interference mitigation by exploiting the inherent sparsity of C-RAN.
  Hao \emph{et al.} \cite{hao2018price-based} considered the mitigation of intra-cluster
 interference and inter-tier interference by exploiting coordinated multipoint
 transmission and by allocating distinct bandwidths to BBUs and RRUs, respectively.

 Fronthaul link selection and user association, which are typically investigated
 together with precoding \cite{ren2017joint,luong2017optimal}, maximize the
 achievable sum-rate of a C-RAN,  given a fixed fronthaul capacity constraint.
 Furthermore, Pan \emph{et al.} \cite{pan2017joint} studied the joint optimization
 of RRU selection, user association and beamforming in the presence of imperfect
 channel state information (CSI). Resource allocation and optimization
 \cite{peng2017cost,liu2015joint,peng2015fronthaul,hao2018small} also offer an
 effective means of enhancing the achievable throughput, given a limited fronthaul
 capacity.

 Upgrading the network infrastructure by replacing copper cabling with fiber cabling
 for fronthaul connections has also been considered as an alternative solution to
 provide high-capacity fronthaul \cite{macho2016next}. However, this approach
 suffers from poor flexibility and high cost in large-scale deployments
 \cite{alimi2018toward,zhang2016fronthauling}. Laying optical fiber to connect the
 RRUs to their host BBU is impossible in city centres of some countries
 \cite{jaber20165Gbackhaul}. Thus wireless backhauling/fronthauling
 \cite{siddique2015wireless,saidulhuq2016backhauling,jaber2018wireless,hu2018joint,Imran_etal2017},
 relying on massive MIMO \cite{marzetta2010noncooperative} and mmWave
 \cite{xiao2017millimeter,Imran_etal2017}, has recently emerged as a promising
 solution for 5G networks due to its flexibility and cost-efficiency 
 \cite{park2015large,zhang2016fronthauling,parsaeefard2017dynamic}. Park \emph{et al.}
 \cite{park2015large} proposed a partially centralized C-RAN based on massive MIMO
 schemes, while Parsaeefard \emph{et al.} \cite{parsaeefard2017dynamic} allocated
 resources by appropriately adjusting the parameters of the RRU, BBU and fronthaul as
 well as the power allocated to UDs.

\subsection{Our Contributions}\label{S1.2}

 A UD's achievable throughput inherently depends on both the UD-to-RRU links and the
 RRU-to-BBU links. Thus all the UD-to-RRU links and RRU-to-BBU links should be
 considered together in order to maximize a C-RAN's capacity. By exploiting the C-RAN's
 superior centralized signal processing/resource control capability over both the
 UD-to-RRU links and the RRU-to-BBU links, we propose to boost the fronthaul capacity
 by globally optimizing the power sharing for both the RRUs and UDs located within a
 BBU's service coverage. Intuitively, allocating more power to channel estimation will
 result in more accurate channel estimates, which increases the achievable throughput.
 But increasing the power allocated to pilot training will reduce the power allocated
 to data transmission, which reduces the achievable data throughput. This paper
 addresses for the first time how to optimize the powers allocated to pilot training
 and to data transmission both for the UDs and RRUs. The main contributions of this
 paper are as follows.
\begin{itemize}
\item [1)] We investigate the C-RAN's configuration, which employs large numbers of
 antennas both at the RRUs and their host BBU. We formulate the ultimately achievable
 uplink sum-rate of the C-RAN as a function of the signal-to-interference-plus-noise
 ratio (SINR) by considering both the UD-to-RRU links and the RRU-to-BBU links.
 Furthermore, we derive a closed-form expression of the asymptotic achievable uplink
 sum-rate in the presence of realistic channel estimation errors both for the UD-to-RRU
 links and the RRU-to-BBU links.
\item [2)] We propose to boost the uplink fronthaul capacity by globally optimizing
 the power sharing between the pilots and data transmission both for RRUs and for the
 UDs within a BBU's service coverage. Specifically, given the power of a UD, a UD's
 power sharing factor controls the specific fractions of power allocated to the UD's
 uplink pilot and to the UD's uplink data transmission, respectively. Similarly, given
 the power of a RRU, the power allocated to the RRU's uplink pilot and the power
 allocated to the RRU's uplink data forwarding are controlled by the RRU's power sharing
 factor. We formulate the uplink sum-rate as a function of the all the UDs' power sharing
 factors and all the RRUs' power sharing factors. We then maximize this uplink sum-rate by
 invoking the global optimization algorithm of the differential evolution algorithm (DEA).  
\end{itemize}

\begin{figure*}[bp!]
\vspace*{-4mm}
\begin{center}
 \includegraphics[width=0.74\textwidth]{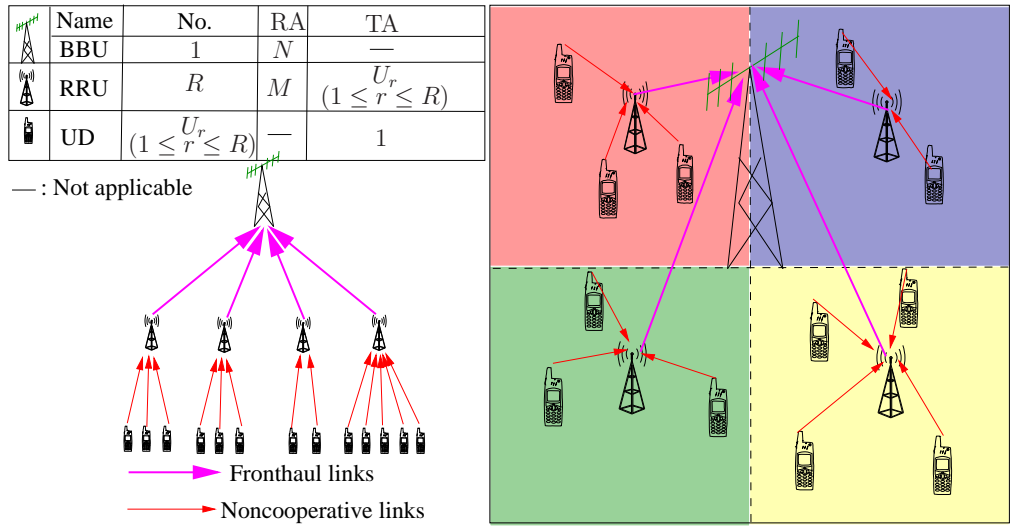}
\end{center}
\vspace*{-5mm}
\caption{An ultra-dense C-RAN architecture for uplink transmission.}
\label{FIG1}
\vspace*{-1mm}
\end{figure*}

\subsection{Notations}\label{S1.3}

 Throughout our discussions, $(\cdot )^{\rm T}$ and $(\cdot )^{\rm H}$ stand for the
 transpose and Hermitian transpose of vector/matrix, respectively. $\bm{0}_{N\times 1}$
 is the $N\times 1$ zero vector, which is abbreviated as $\bm{0}_{N}$, and 
 $\bm{0}_{N\times N }$ denotes the $N\times N$ zero matrix, while following the convention,
 the $N \times N$ identity matrix is represented by $\bm{I}_N$. The the expectation
 operation is represented by $\mathcal{E}\{\cdot \}$ and $\bm{vec}(\bm{A})$ denotes the
 column stacking operation applied to matrix $\bm{A}$. The diagonal matrix with
 $a_1,\cdots ,a_M$ at its diagonal entries is denoted by $\text{diag}\{a_1,\cdots ,
 a_M\}$ and the block diagonal matrix $\text{Bdiag}\{\bm{B}_1, \cdots , \bm{B}_M\}$ has 
 $\bm{B}_1, \cdots , \bm{B}_M$ as its block diagonal entries, while $ \otimes $ denotes
 the Kronecker product and $\text{Tr}(\cdot )$ is the trace operator. Furthermore,
 $\bm{A}_{[i:\,]}$ and $\bm{A}_{[\,:i]}$ denote the $i$th row and $i$th column of $\bm{A}$,
 respectively. Furthermore, the subscripts $\text{tx}$ and $\text{rx}$ indicate that the
 variable considered is at the transmitter and the variable considered is at the receiver,
 respectively. The superscript $(r/u)$ denotes a variable between the UDs and the RRU, and
 the superscript $(b/r)$ denotes a variable between RRUs and the BBU, respectively. The
 superscripts $(u,d)$ and $(u,p)$ are variables related with UD's data transmission and
 pilot training, and the superscripts $(r,d)$ and $(r,p)$ are variables related with RRU's
 data transmission and pilot training, respectively.

 For easy reference, below we list the key mathematical symbols used in the manuscript.
\begin{itemize}[leftmargin=0.6in]
\item[$K$:] Total number of UDs.
\item[$R$:] Number of RRUs.
\item[$U_r$:] Number of UDs served by the $r$th RRU.
\item[$M$:] Number of RAs equipped by RRU.
\item[$N$:] Number of RAs equipped by BBU.
\item[$P_{\rm UD}$:] Total power of an UD.
\item[$P_{\rm RRU}$:] Total power of a RRU.
\item[$\rho_{r}$:] Receiver's efficiency factor of the $r$th RRU.
\item[$P_{{\rm sp},r}$:] Received signal processing power of the $r$th RRU.
\item[$L_{\text{\rm pathloss}}$:] Pathloss.
\item[$d_{u_r}$:] Distance between the $u_r$th UD and the $r$th RRU.
\item[$d_r$:] Distance between the $r$th RRU and its host BBU.
\item[$\lambda_r^{(r/u)}$:] Normalization factor at the $r$th RRU.
\item[$\lambda^{(b/r)}$:] Normalization factor at the BBU.
\item[$\eta_{u_r}^{(r/u)}$:] Power sharing factor of the $u_r$th UD.
\item[$\eta_{u_r}^{(b/r)}$:] Power sharing factor of the TA for forwarding the $u_r$th UD's data.
\item[$P_{{\rm tx},u_r}^{(u,x)}$:] Transmit power of the $u_r$th UD, $x=p$ for pilot
 and $x=d$ for data.
\item[$P_{{\rm tx},r(u_r)}^{(r,x)}$:] Transmit power of the $u_r$th TA at the $r$th RRU,
 $x=p$ for pilot and $x=d$ for data.
\item[$P_{{\rm rx},u_r}^{(u,x)}$:] Receive signal power of the $u_r$th UD at the $r$th RRU,
 $x=p$ for pilot and $x=d$ for data.
\item[$\bm{P}_{{\rm rx},r/r'}^{(u,x)}$:] Receive signal powers of the $r'$th RRU's UDs
 at the $r$th RRU, $x=p$ for pilot and $x=d$ for data.
\item[$\bm{P}_{\rm rx}^{(r,d)}$:] Receive signal powers at the BBU.
\item[$\bm{s}_r$:] Transmit signals of the UDs served by the $r$th RRU.
\item[$\bm{H}_r^{(r/u)}$:] Uplink MIMO channel between  the $r$th RRU and its UDs.
\item[$\bm{H}_{r/r'}^{(r/u)}$:] Interfering channel between the $r'$th RRU's UDs and the $r$th RRU.
\item[$\widetilde{\bm{y}}_r$:] Signals received at the $r$th RRU.
\item[$\widetilde{\bm{n}}_r$:] AWGN vector at the $r$th RRU.
\item[$\bm{W}_r^{(r/u)}$:] Receiver combining matrix used by the $r$th RRU.
\item[$\bm{y}_r$:] Uplink signals after combining at the $r$th RRU.
\item[$\bm{n}_r$:] Noise output after the $r$th RRU's combining.
\item[$ \bm{A}_r$:] Power amplification at the $r$th RRU.
\item[$\bar{\bm{y}}$:] Transmit signals consisting of all the signals
 transmitted by all the $R$ RRUs.
\item[$K_{Rice}$:] Rician factor.
\item[$\bm{H}^{(b/r)}$:] MIMO Rician channel between the $K$ forwarding TAs of all
 the $R$ RRUs and their host BBU.
\item[$\bm{H}_{\rm d}^{(b/r)}$:] Deterministic part of the Rician channel.
\item[$\bm{H}_{\rm r}^{(b/r)}$:] Scattered component of the Rician channel .
\item[$\breve{\bm{y}}$:] Signals received at the BBU.
\item[$\breve{\bm{n}}$:] AWGN vector at the BBU.
\item[$\bm{W}^{(b/r)}$:] BBU's combining matrix.
\item[$\check{\bm{n}}$:] Noise output after BBU's combining.
\item[$\check{\bm{y}}$:] Uplink signals after combining at BBU.
\end{itemize}

 Additionally, the main abbreviations used are listed below.
\begin{itemize}[leftmargin=0.6in]
\item[AWGN:] Additive white Gaussian noise
\item[BBU:] Baseband unit
\item[C-RAN:] Centralized radio access network
\item[CSI:] Channel state information
\item[DEA:] Differential evolution algorithm
\item[i.i.d.:] Independently and identically distributed
\item[LoS:] Line-of-sign
\item[MF:] Matched filter
\item[mmWave:] Millimeter-wave
\item[RA:] Receive antenna
\item[RRU:] Remote radio unit
\item[SINR:] Signal to interference plus noise ratio
\item[TA:] Transmit antenna
\item[UD:] User device
\end{itemize}

 The rest of this paper is organized as follows. Section~\ref{S2} describes the
 massive MIMO aided uplink C-RAN architecture, and Section~\ref{S3} derives its
 closed-form achievable asymptotic uplink sum-rate. The global optimization
 metric as a function of all the power sharing factors and how to maximize it
 are presented in Section~\ref{S4}. Our simulation results are presented in
 Section~\ref{S5} for demonstrating the efficiency of the proposed approach, whilst
 our conclusions are given in Section~\ref{S6}. 

\section{System Model}\label{S2}

 Consider an uplink C-RAN architecture as illustrated in Fig.~\ref{FIG1}. Each RRU is
 connected to its host BBU via a wireless fronthaul link. Assume that the frequency
 is reused within the coverage of a BBU, while orthogonal access is adopted by the 
 different BBUs. Thus there is no interference between the UDs located in the different
 BBUs' coverage areas. Hence we only have to consider a single BBU's coverage. The BBU
 employs $N$ receive antennas (RAs) to serve $R$ RRUs, while each RRU is equipped with
 $M$ RAs. The $r$th RRU employs $U_r$ transmit antennas (TAs) and serves $U_r$ single-TA
 UDs, where $1\le r\le R$. The total number of UDs within the BBU's coverage is $K=
 \sum\nolimits_{r=1}^R U_r$.  According to the well-known spatial multiplexing gaining or
 spatial degree of freedom, the number of independent data streams supported cannot be
 higher than the number of receive antennas. The spatial degree of freedom between the
 $r$th RRU and its supporting UDs is given by $\min\{U_r, M\}$
 \cite{zhang2018adaptive,zhang2018regularized}. Therefore, the number of UDs served by
 the $r$th RRU is no more than the number of the $r$th RRU's RAs, i.e., $U_r\le M$. We
 further assume that $M\gg U_r$ for $r\in\{1,2,\cdots,R\}$. Since the $r$th RRU uses
 $U_r$ TAs for forwarding its serving UDs' data to the BBU, the total number of TAs for
 all the $R$ RRUs is $K$, which equals to the number of uplink fronthaul streams. Clearly,
 $K$ is no more than the number of the BBU's RAs, i.e., $K \le N$. 

\subsection{Power Consumption Preliminaries}\label{S2.1}

 For a fair power allocation, all the UDs have the same total power $P_{\rm UD}$, which
 is shared by pilot training and data transmission of each UD. Explicitly, for the 
 link of the $u_r$th UD to the $r$th RRU, the power allocation between pilot training
 and data transmission is controlled by a power sharing factor $0 < \eta_{u_r}^{(r/u)}
 < 1$. Let the power of pilot training be $P_{{\rm tx},u_r}^{(u,p)}$ and that of 
 data transmission be $P_{{\rm tx},u_r}^{(u,d)}$, respectively, for the $u_r$th UD served
 by the $r$th RRU.  Then,
\begin{align}
 P_{{\rm tx},u_r}^{(u,p)} =& \eta_{u_r}^{(r/u)} P_{\rm UD} , \label{eq1} \\
 P_{{\rm tx},u_r}^{(u,d)} =& \big(1 - \eta_{u_r}^{(r/u)}\big) P_{\rm UD} . \label{eq2}
\end{align}
 For convenience, we introduce the new symbol $x$ with $x=p$ indicating pilot training
 and  $x=d$ indicating data transmission. At the $r$th RRU, the received signal power
 $P_{{\rm rx},u_r}^{(u,x)}$ of the $u_r$th UD is given by
\begin{align}\label{eq3}
 P_{{\rm rx},u_r}^{(u,x)} =& P_{{\rm tx},u_r}^{(u,x)} 10^{ - L_{\rm pathloss}(u_r) / 10} ,
\end{align}
 where the pathloss $L_{\rm pathloss}(u_r)$ is given by \cite{parsons2000the}
\begin{align}\label{eq4}
L_{\text{\rm pathloss}}(u_{r}) \,[\text{dB}] \!\! =\!\! -154 \! +\! 20\log_{10}\big(f_{c}\big)
 \!+\! 20\log_{10}\big(d_{u_r}\big) ,
\end{align}
 in which $d_{u_r}$ [m] is the distance between the $u_r$th UD and its host, the
 $r$th RRU, and $f_c$ [Hz] is the carrier frequency. For the 5G system, $f_c=3.4$\,GHz has been allocated in United
 Kingdom (UK) \cite{ofcom2017update}, which will
 be considered in our investigations.

 Each  RRU is allocated  with the same total power $P_{\rm RRU}$ for the sake of
 fairness, which is shared by the received signal processing of detecting the served UDs'
 data as well as pilots and data transmission for forwarding these UDs' data to the BBU.
 Intuitively, the power consumed by the $r$th RRU's received signal processing is related
 to the uplink sum-rate of its serving UDs, which is dominated by the uplink transmit
 power. Since the accurate modeling of this received signal processing is absent in the
 literature, we approximately model the $r$th RRU's received signal processing power
 consumption $P_{{\rm sp},r}$ as a function of its serving UDs' uplink transmit power by
\begin{align}\label{eq5}
 P_{{\rm sp},r} =& \rho_r U_r P_{\rm UD} ,
\end{align}
 where $\rho_{r}$ is the receiver's efficiency factor of the $r$th RRU. The remaining power
 $\big(P_{\rm RRU}-P_{{\rm sp},r}\big)$ of the $r$th RRU is shared by its $U_r$ TAs for 
 pilots and data transmission. Let $\eta_{u_r}^{(b/r)}$ be the power sharing factor of
 the TA for forwarding the $u_r$th UD's data.  Then, the pilot training power
 $P_{{\rm tx},r(u_r)}^{(r,p)}$ and data transmission power $P_{{\rm tx},r(u_r)}^{(r,d)}$ for
 the  $u_r$th TA are given by
\begin{align}
 P_{{\rm tx},r(u_r)}^{(r,p)} =& \eta_{u_r}^{(b/r)} {\big(P_{\rm RRU} - P_{{\rm sp},r}\big)}\big/
  {U_r} , \label{eq6} \\
 P_{{\rm tx},r(u_r)}^{(r,d)} =& \big(1 - \eta_{u_r}^{(b/r)}\big) {\big(P_{\rm RRU} -
  P_{{\rm sp},r}\big)}\big/ {U_r} . \label{eq7}
\end{align}
 At the BBU, the received signal power $P_{{\rm rx},r(u_r)}^{(r,x)}$ is related to the
 transmit signal power $P_{{\rm tx},r(u_r)}^{(r,x)}$ by a similar pathloss model 
 $P_{{\rm rx},r(u_r)}^{(r,x)}=P_{{\rm tx},r(u_r)}^{(r,x)} 10^{- L_{\rm pathloss}\big[r(u_r)\big] / 10}$.
 Since mmWave communication is established between the RRUs and their host
 BBU, the pathloss $L_{\rm pathloss}\big[r(u_r)\big]$ of the RRU-to-BBU link is given by
 \cite{METIS2020,rappaport2017overview}
\begin{align}\label{eq8}
L_{\rm pathloss}\big[r(u_r)\big] \,[\text{dB}] \!\!= \!\!3.34 \!+\! 18.62\log_{10}(f_{mm})  \!+ \!22\log_{10}(d_r)  ,
\end{align}
 where $d_r$\,[m] is the distance between the $r$th RRU and its host BBU, while $f_{mm}$\,[GHz]
 is the carrier frequency. For the mmWave based 5G system in the UK, $f_{mm}=26$\,GHz is allocated
 \cite{ofcom2017update}, which is used for our investigations.

\begin{figure*}[tp!]\setcounter{equation}{12}
\vspace*{-5mm}
\begin{align}\label{eq13}
 \bm{vec}\left(\widehat{\bm{H}}_r^{(r/u)}\right)  =& \bm{\Psi}_{\bm{H}_r^{(r/u)}}
  \Big(\sigma_{\widetilde{n}}^2\big(\bm{P}_{{\rm rx},r}^{(u,p)}\big)^{-1} \otimes \bm{I}_M
  + \bm{\Psi}_{\bm{H}_r^{(r/u)}}\Big)^{-1} \bm{vec}\big( \widetilde{\bm{Y}}_r\bm{X}_r^{(p)}\big) ,
\end{align}
\hrulefill
\vspace*{-5mm}
\end{figure*}

\subsection{Signal Model of UD-to-RRU}\label{S2.2}

 Obviously the pilot training and data transmission have the same signal model. Again,
 we introduce the symbol $x$, with $x=p$ indicating the pilot training and $x=d$
 representing data transmission, respectively. Then, the signals received at the $r$th RRU
 $\widetilde{\bm{y}}_r\in \mathbb{C}^M$ can be expressed generically as\setcounter{equation}{8}
\begin{align}\label{eq9}
 \widetilde{\bm{y}}_r =& \bm{H}_r^{(r/u)}\big(\bm{P}_{{\rm rx},r}^{(u,x)}\big)^{\frac{1}{2}}
  \bm{s}_{r} + \!\!\!\!\! \sum\limits_{r'=1, r'\neq r}^R \!\!\!\!\! \bm{H}_{r/r'}^{(r/u)}
  \big(\bm{P}_{{\rm rx},r/r'}^{(u,x)}\big)^{\frac{1}{2}} \bm{s}_{r'}\! +\! \widetilde{\bm{n}}_r ,
\end{align}
 where $\bm{H}_r^{(r/u)}\in \mathbb{C}^{M\times U_r}$ is the uplink MIMO channel between
 the $r$th RRU and its UDs, $\bm{s}_r=\big[s_1^{(r)} ~ s_2^{(r)} \cdots s_{U_r}^{(r)}\big]^{\rm T}$
 is the transmit signal vector with $\mathcal{E}\left\{|s_{u_r}^{(r)}|\right\}=1$ for
 $u_r \in \{1,2,\cdots,U_r\}$,  and $\widetilde{\bm{n}}_r\in \mathbb{C}^M$ is the
 additive white Gaussian noise (AWGN) vector with the distribution
 $\mathcal{CN}(\bm{0}_M,\sigma_{\widetilde{n}}^2\bm{I}_M)$. Still referring to (\ref{eq9}), 
 and $\bm{P}_{{\rm rx},r}^{(u,x)}=\text{diag}\big\{P_{{\rm rx},r(1)}^{(u,x)}, \cdots ,
 P_{{\rm rx},r(U_r)}^{(u,x)}\big\}$ are the received powers of the $U_r$ UDs' signals at the
 $r$th RRU, while $\bm{H}_{r/r'}^{(r/u)}$ is the interfering channel matrix between
 the $r'$th RRU's UDs and the $r$th RRU, and $\bm{P}_{{\rm rx},r/r'}^{(u,x)}=\text{diag}\big\{
 P_{{\rm rx},r/r'(1)}^{(u,x)}, \cdots , P_{{\rm rx},r/r'(U_{r'})}^{(u,x)}\big\}$ are the received
 powers of the $r'$th RRU's UDs at the $r$th RRU. Since all the UDs' signals suffer from the
 same noise at the RRU, the noise power at all the UD-to-RRU links are identical. Furthermore,
 the second term in (\ref{eq9}) represents the interference imposed by the UDs of the
 adjacent RRUs.
 
 Because the UDs are randomly distributed in the RRUs' coverage areas and there are many
 obstructions between the UDs and their host RRU, the direct line-of-sight (LoS) paths may
 always be blocked. Hence, the channels between the UDs and their host RRU are Rayleigh
 channels, and the UD-to-RRU MIMO channel matrix can be expressed as
\begin{align}\label{eq10}
 \bm{H}_{r}^{(r/u)} =& \left(\bm{R}_{{\rm rx},r}^{(r/u)}\right)^{\frac{1}{2}} \bm{G}_{r}^{(r/u)}
  \left(\bm{R}_{{\rm tx},u}^{(r/u)}\right)^{\frac{1}{2}} ,
\end{align}
 where $\bm{R}_{{\rm rx},r}^{(r/u)}\in \mathbb{C}^{M\times M}$ is the spatial correlation
 matrix of the $r$th RRU's $M$ RAs and $\bm{R}_{{\rm tx},u}^{(r/u)}\in\mathbb{C}^{U_r\times U_r}$
 is the spatial correlation matrix of the $U_r$ UDs, while $\bm{G}_{r}^{(r/u)}\in 
 \mathbb{C}^{M\times U_r}$ has the independently identically distributed (i.i.d.)
 complex entries and each of them has the distribution of $\mathcal{CN}(0,1)$. Because
 the UDs are randomly distributed and they are independent of each other, there is no
 correlation between the TAs of different UDs and we have $\bm{R}_{{\rm tx},u}^{(r/u)}=
\bm{I}_{U_r}$. 

 Let the receiver combining matrix used by the $r$th RRU be $\bm{W}_r^{(r/u)}\in
 \mathbb{C}^{U_r\times M}$. The uplink signals $\bm{y}_r\in \mathbb{C}^{U_r}$ after
 combining at the $r$th RRU can be expressed as
\begin{align}\label{eq11}
 \bm{y}_r &= \sqrt{\lambda_r^{(r/u)}}\bm{W}_r^{(r/u)} \bm{H}_r^{(r/u)}
  \big(\bm{P}_{{\rm rx},r}^{(u,x)}\big)^{\frac{1}{2}} \bm{s}_r \nonumber \\
 &  +\! \sqrt{\lambda_r^{(r/u)}} \bm{W}_r^{(r/u)} \!\!\!\!\! \sum\limits_{r'=1, r'\neq r}^R \!\!\!\!\!
  \bm{H}_{r/r'}^{(r/u)} \big(\bm{P}_{{\rm rx},r/r'}^{(u,x)}\big)^{\frac{1}{2}} \bm{s}_{r'}\! +\! \bm{n}_r , \!
\end{align}
 where $\lambda_r^{(r/u)}\! =\! 1\big/ \Big(\frac{1}{U_r}\mathcal{E}\Big\{
 \text{Tr}\big\{\bm{W}_r^{(r/u)}\big(\bm{W}_r^{(r/u)}\big)^{\rm H}\big\}\Big\}\Big)$ is
 the normalization factor, and $\bm{n}_r=\sqrt{\lambda_r^{(r/u)}}\bm{W}_r^{(r/u)}
 \widetilde{\bm{n}}_r\in \mathbb{C}^{U_r}$ is the effective noise vector having the 
 distribution of ${\cal CN}\big(\bm{0}_{U_r},\bm{\Sigma}_{\bm{n}_r}\big)$ with the covariance
 matrix $\bm{\Sigma}_{\bm{n}_r}=\lambda_r^{(r/u)} \sigma_{\widetilde{n}}^2\bm{W}_r^{(r/u)}
 \big(\bm{W}_r^{(r/u)}\big)^{\rm H}$. When a matched-filter (MF) is used for  uplink
 combining, we have
\begin{align}\label{eq12}
 \bm{W}_r^{(r/u)} =& \left(\widehat{\bm{H}}_r^{(r/u)}\right)^{\rm H} ,
\end{align}
 where $\widehat{\bm{H}}_r^{(r/u)}$ is the estimate of the uplink channel $\bm{H}_r^{(r/u)}$.
 
 The optimal minimum mean square error (MMSE) channel estimator \cite{kay2003fundamentals}
 is given by (\ref{eq13}) at the top of this page\footnote{We assume that the pilot
 contamination imposed by the UDs served by the adjacent RRUs has been eliminated by optimal
 pilot design \cite{zhang2014pilot,guo2016optimal}.}, where $\widetilde{\bm{Y}}_r\in
 \mathbb{C}^{M\times U_r}$ is the received signal matrix over the $U_r$ pilot symbols, and
 $\bm{X}_r^{(p)}\in \mathbb{C}^{U_r\times U_r}$ is the pilot symbol matrix with $\bm{X}_r^{(p)}
 \big(\bm{X}_r^{(p)}\big)^{\rm H}=\bm{I}_{U_r}$, while $\bm{P}_{{\rm rx},r}^{(u,p)}=
 \text{diag}\Big\{P_{{\rm rx},r(1)}^{(u,p)},\cdots ,P_{{\rm rx},r(U_r)}^{(u,p)}\Big\}$ are
 the received powers of the pilot symbols. Furthermore, in (\ref{eq13}),
 $\bm{\Psi}_{\bm{H}_r^{(r/u)}}$ denotes the covariance matrix of $\bm{vec}\big(\bm{H}_r^{(r/u)}\big)$,
 which is given by\setcounter{equation}{13}
\begin{align}\label{eq14}
 \bm{\Psi}_{\bm{H}_r^{(r/u)}} =& \mathcal{E}\left\{\bm{vec}\left(\bm{H}_r^{(r/u)}\right)
  \bm{vec}\left(\bm{H}_r^{(r/u)}\right)^{\rm H}\right\} \nonumber \\
 =& \bm{I}_{U_r} \otimes \bm{R}_{{\rm rx},r}^{(r/u)} \in\mathbb{C}^{M U_r\times M U_r}.
\end{align}
 The MMSE estimate (\ref{eq13}) follows the distribution \cite{kay2003fundamentals}
\begin{align}\label{eq15}
 \bm{vec}\left(\widehat{\bm{H}}_r^{(r/u)}\right)\sim \mathcal{CN}\left(\bm{0}_{M U_r},
  \bm{\Psi}_{\widehat{\bm{H}}_r^{(r/u)}}\right) ,
\end{align}
 with the covariance matrix $\bm{\Psi}_{\widehat{\bm{H}}_r^{(r/u)}}$ given by
\begin{align}\label{eq16}
 \bm{\Psi}_{\widehat{\bm{H}}_r^{(r/u)}} =& \bm{\Psi}_{\bm{H}_r^{(r/u)}}\Big(\sigma_{\widetilde{n}}^2
  \big(\bm{P}_{{\rm rx},r}^{(u,p)}\big)^{-1} \otimes \bm{I}_M + \bm{\Psi}_{\bm{H}_r^{(r/u)}}\Big)^{-1} \nonumber \\
 & \times \bm{\Psi}_{\bm{H}_r^{(r/u)}} \in\mathbb{C}^{M U_r\times M U_r} .
\end{align}
 The relationship between the MMSE channel estimate $\bm{vec}\big(\widehat{\bm{H}}_r^{(r/u)}\big)$
 and the true channel $\bm{vec}\big(\bm{H}_r^{(r/u)}\big)$ is given by\setcounter{equation}{16}
\begin{align}\label{eq17}
 \bm{vec}\left(\bm{H}_r^{(r/u)}\right) =& \bm{vec}\left(\widehat{\bm{H}}_r^{(r/u)}\right)
 + \bm{vec}\left(\widetilde{\bm{H}}_r^{(r/u)}\right) , 
\end{align}
 where the channel estimation error $\bm{vec}\big(\widetilde{\bm{H}}_r^{(r/u)}\big)$ is
 statistically independent of both $\bm{vec}\big(\widehat{\bm{H}}_r^{(r/u)}\big)$ and
 $\bm{vec}\big(\bm{H}_r^{(r/u)}\big)$. Moreover, the distribution of
 $\bm{vec}\big(\widetilde{\bm{H}}_r^{(r/u)}\big)$ is
\begin{align}\label{eq18}
 \bm{vec}\left(\widetilde{\bm{H}}_r^{(r/u)}\right)\sim & \mathcal{CN}\left(\bm{0}_{M U_r},
  \bm{\Psi}_{\widetilde{\bm{H}}_r^{(r/u)}}\right) ,
\end{align}
 with the covariance matrix $\bm{\Psi}_{\widetilde{\bm{H}}_r^{(r/u)}}$ given by
\begin{align}\label{eq19}
 \bm{\Psi}_{\widetilde{\bm{H}}_r^{(r/u)}} =& \bm{\Psi}_{\bm{H}_r^{(r/u)}} -
  \bm{\Psi}_{\widehat{\bm{H}}_r^{(r/u)}} .
\end{align}

\subsection{Signal Model of RRU-to-BBU}\label{S2.3}

 The UDs' data signals received by their host RRUs are forwarded to the BBU after power
 amplification. Ideally, the $u_r$th UD's data received by the $r$th RRU is scaled
 to its transmit signal constellation by a power amplification coefficient $a_{u_r}^{(r)}
 =\frac{1}{P_{{\rm rx},r(u_r)}^{(u,d)}}$. Hence, the power amplification at the $r$th RRU
 is represented by the diagonal matrix given by\setcounter{equation}{19}
\begin{align}\label{eq20}
 \bm{A}_r =& \text{diag}\big\{a_1^{(r)},\cdots a_{U_r}^{(r)}\big\} = \left(\bm{P}_{{\rm rx},r}^{(u,d)}\right)^{-1} ,
\end{align}
 and the $r$th RRU forwards the amplified signal $\bar{\bm{y}}_r=\bm{A}_r^{\frac{1}{2}}
 \bm{y}_r\in\mathbb{C}^{U_r}$ to its host BBU via the fronthaul links. Explicitly, let
 $\bar{\bm{y}}\in \mathbb{C}^K$ be the transmit signal vector consisting of all the signals
 transmitted by the $R$ RRUs, which is given by
\begin{align}\label{eq21}
 \bar{\bm{y}} =& \left[\bar{\bm{y}}_1^{\rm T} \cdots \bar{\bm{y}}_R^{\rm T}\right]^{\rm T} 
 \!\! =\! \bm{A}^{\frac{1}{2}} \big(\bm{\lambda}^{(r/u)}\big)^{\frac{1}{2}} \bm{W}^{(r/u)} \bm{H}^{(r/u)}
  \big(\bm{P}_{\rm rx}^{(u,d)}\big)^{\frac{1}{2}} \bm{s} \nonumber \\ &
  + \bm{A}^{\frac{1}{2}} \big(\bm{\lambda}^{(r/u)}\big)^{\frac{1}{2}} \bm{W}^{(r/u)} 
  \sum\limits_{r'=2}^R \bm{V}^{(r/u)}_{r'} \widetilde{\bm{s}}_{r'}\! +\! \bm{A}^{\frac{1}{2}} \bm{n} ,
\end{align}

\begin{figure*}[tp!]\setcounter{equation}{35}
\vspace*{-5mm}
\begin{align}\label{eq36}
 \bm{vec}\big(\widehat{\bm{H}}^{(b/r)}\big)\! =& \nu \bm{vec}\big(\bm{H}_{\rm d}^{(b/r)}\big)\! +\! \zeta^{2}
  \bm{\Psi}_{\bm{H}_{\rm r}^{(b/r)}} \Big(\sigma_{\breve{n}}^2 \big(\bm{P}_{{\rm rx}}^{(r,p)}\big)^{-1}
  \! \otimes\! \bm{I}_N\! +\! \zeta^{2}\bm{\Psi}_{\bm{H}_{\rm r}^{(b/r)}}\Big)^{-1}\! 
  \bm{vec}\big( \breve{\bm{Y}}\bm{X}^{(p)}_b\big) ,
\end{align}
\hrulefill
\vspace*{-5mm}
\end{figure*}

\begin{figure*}[bp!]\setcounter{equation}{45}
\vspace*{-5mm}
\hrulefill
\begin{align} \label{eq46}
 \check{y}_{k^*} =& \sqrt{a_{k^*} \lambda^{(b/r)} \lambda_{k^*}^{(r/u)} P_{{\rm rx},k^*}^{(r,d)}
  P_{{\rm rx},k^*}^{(u,d)}} \bm{W}_{[k^*:\,]}^{(b/r)} \bm{H}_{[\,:k^*]}^{(b/r)}
  \bm{W}_{[k^*:\,]}^{(r/u)} \bm{H}_{[\,:k^*]}^{(r/u)} s_{k^*} \nonumber \\
 & +\sum\limits_{k=1,k\neq k^*}^K  \sqrt{a_{k^*} \lambda^{(b/r)} \lambda_{k^*}^{(r/u)}
  P_{{\rm rx},k}^{(r,d)} P_{{\rm rx},k}^{(u,d)}} \bm{W}_{[k^*:\,]}^{(b/r)} \bm{H}_{[\,:k]}^{(b/r)}
  \bm{W}_{[k:\,]}^{(r/u)} \bm{H}_{[\,:k]}^{(r/u)} s_k \nonumber \\
 & +\sum\limits_{k=1}^K \sum\limits_{j=1,j \neq k}^K \delta\big(\Delta_{j,k}\big)
  \sqrt{a_{k^*} \lambda^{(b/r)} \lambda_k^{(r/u)} P_{{\rm rx},k}^{(r,d)}} \bm{W}_{[k^*:\,]}^{(b/r)}
  \bm{H}_{[\,:k]}^{(b/r)} \bm{W}_{[k:\,]}^{(r/u)} \sum\limits_{r'=2}^R \bm{V}^{(r/u)}_{r' ~~ [\,:j]}
  \widetilde{s}_{r',j} + \check{\bar{n}}_{k^*} ,
\end{align}
\hrulefill
\begin{align}\label{eq47}
 \bar{P}_{{\rm S},k^*} =& a_{k^*} \lambda^{(b/r)} \lambda_{k^*}^{(r/u)} P_{{\rm rx},k^*}^{(r,d)} P_{{\rm rx},k^*}^{(u,d)}
  \mathcal{E}\left\{ \Big(\widehat{\bm{H}}_{[\,:k^*]}^{(b/r)}\Big)^{\rm H} \widehat{\bm{H}}_{[\,:k^*]}^{(b/r)}
  \Big(\widehat{\bm{H}}_{[\,:k^*]}^{(r/u)}\Big)^{\rm H} \bm{H}_{[\,:k^*]}^{(r/u)}
  \Big(\bm{H}_{[\,:k^*]}^{(r/u)}\Big)^{\rm H} \widehat{\bm{H}}_{[\,:k^*]}^{(r/u)}
  \Big(\widehat{\bm{H}}_{[\,:k^*]}^{(b/r)}\Big)^{\rm H} \widehat{\bm{H}}_{[k^*:\,]}^{(b/r)}\right\} .
\end{align}
\vspace*{-2mm}
\end{figure*}
 where we have \setcounter{equation}{21}
\begin{align}
 \bm{A} =& \text{Bdiag}\left\{\bm{A}_1,\cdots ,\bm{A}_R\right\}\in \mathbb{C}^{K\times K} , \label{eq22} \\
 \bm{\lambda}^{(r/u)} =& \text{Bdiag}\Big\{\lambda_1^{(r/u)}\bm{I}_{U_1},\cdots ,
  \lambda_R^{(r/u)}\bm{I}_{U_R}\Big\}\in\! \mathbb{C}^{K\times K}\! ,\! \label{eq23} \\
 \bm{W}^{(r/u)} =& \text{Bdiag}\left\{\bm{W}_1^{(r/u)},\cdots ,\bm{W}_R^{(r/u)}\right\}
  \in\! \mathbb{C}^{K\times MR} , \label{eq24} \\
 \bm{H}^{(r/u)} =& \text{Bdiag}\left\{\bm{H}_1^{(r/u)},\cdots ,\bm{H}_R^{(r/u)}\right\}
  \in\! \mathbb{C}^{MR\times K} , \label{eq25} 
\end{align}
\begin{align}
 \bm{P}_{\rm rx}^{(u,d)} =& \text{Bdiag}\left\{\bm{P}_{{\rm rx},1}^{(u,d)},\cdots ,
  \bm{P}_{{\rm rx},R}^{(u,d)}\right\}\in\! \mathbb{C}^{K\times K} , \label{eq26} \\
 \bm{s} =& \left[\bm{s}_1^{\rm T}\cdots \bm{s}_R^{\rm T}\right]^{\rm T}\in\! \mathbb{C}^K , \label{eq27} \\
 \bm{n} =& \left[\bm{n}_1^{\rm T} \cdots \bm{n}_R^{\rm T}\right]^{\rm T}\in \mathbb{C}^K , \label{eq28} \\
 \bm{V}_{r'}^{(r/u)} =& \text{Bdiag}\left\{\bm{V}_{1/(r'\,\text{mod}\,R)}^{(r/u)},\cdots ,
 \bm{V}_{R/((r'+R-1)\,\text{mod}\,R)}^{(r/u)}\right\} \nonumber \\
 & \in \mathbb{C}^{M R \times K} , \label{eq29} \\
 \widetilde{\bm{s}}_{r'} =& \left[\bm{s}_{(r'\,\text{mod}\,R)}^{\rm T},\cdots, 
  \bm{s}_{((r'+R-1)\,\text{mod}\,R)}^{\rm T}\right]^{\rm T}\in \mathbb{C}^K . \label{eq30} 
\end{align}
 In (\ref{eq29}), $\bm{V}_{r/(t\,\text{mod}\,R)}^{(r/u)}\in \mathbb{C}^{M\times U_{t}}$, 
 $t= r',r'+1,\cdots ,(r'+R-1)$, depends on the value of $(t\,\text{mod}\,R)$, which is given by
\begin{align} \label{eq31}
 \bm{V}_{r/(t\,\text{mod}\,R)}^{(r/u)}\! =& \! \left\{ \!\!\! \begin{array}{cl}
  \bm{H}_{r/(t\,\text{mod}\,R)}^{(r/u)} \big(\bm{P}_{{\rm rx},r/(t\,\text{mod}\,R)}^{(u,d)}\big)^{\frac{1}{2}} ,
  \!\! &\!\! t\,\text{mod}\,R \neq 0 , \\
  \bm{H}_{r/R}^{(r/u)} \big(\bm{P}_{{\rm rx},r/R}^{(u,d)}\big)^{\frac{1}{2}} \!\! & \!\! 
  t\,\text{mod}\,R = 0 .
\end{array} \! \right. \! 
\end{align}
 Similarly, in (\ref{eq30}), if $t\,\text{mod}\,R \neq 0$, $\bm{s}_{(t\,\text{mod}\,R)}$ is 
 as it is, while if $t\,\text{mod}\,R = 0$, $\bm{s}_{(t\,\text{mod}\,R)}=\bm{s}_{R}$.

 The signals received at the BBU $\breve{\bm{y}}\in \mathbb{C}^N$ are expressed as
\begin{align}\label{eq32}
 \breve{\bm{y}} =& \bm{H}^{(b/r)}\big(\bm{P}_{\rm rx}^{(r,d)}\big)^{\frac{1}{2}} \bar{\bm{y}}
  + \breve{\bm{n}} ,
\end{align}
 where $\breve{\bm{n}}\sim \mathcal{CN}\big(\bm{0}_N,\sigma_{\breve{n}}^2\bm{I}_N\big)$ is
 the AWGN vector, $\bm{P}_{\rm rx}^{(r,d)}=\text{diag}\Big\{P_{{\rm rx},1}^{(r,d)},
 \cdots ,P_{{\rm rx},K}^{(r,d)}\Big\}$ are the received signal powers at the BBU, and
 $\bm{H}^{(b/r)}\in \mathbb{C}^{N\times K}$ is the MIMO channel matrix between the $K$
 forwarding-TAs of the $R$ RRUs and their host BBU. The RRUs are generally stationary
 and are carefully positioned, so that the direct LoS paths always exist between the
 RRUs and their host BBU. Thus the channels between the RRUs and their host BBU are
 Rician channels and, therefore, the MIMO channel matrix $\bm{H}^{(b/r)}$ is given by
\begin{align}\label{eq33}
 \bm{H}^{(b/r)} =& \nu \bm{H}_{\rm d}^{(b/r)} + \zeta \bm{H}_{\rm r}^{(b/r)} \nonumber \\
 =& \nu\bm{H}_{\rm d}^{(b/r)} + \zeta \left(\bm{R}^{(b/r)}_{{\rm rx},b}\right)^{\frac{1}{2}}
  \bm{G}^{(b/r)}\left(\bm{R}^{(b/r)}_{{\rm tx},r}\right)^{\frac{1}{2}} ,
\end{align}
 where $\bm{H}_{\rm d}^{(b/r)}$ is the deterministic part of the Rician channel satisfying
 $\text{Tr}\Big\{\bm{H}_{\rm d}^{(b/r)}\left(\bm{H}_{\rm d}^{(b/r)}\right)^{\rm H}\Big\}=N K$,
 $\nu=\sqrt{\frac{K_{Rice}}{1+K_{Rice}}}$ and $\zeta=\sqrt{\frac{1}{1+K_{Rice}}}$ with $K_{Rice}$
 being the Rician factor, while $\bm{H}^{(b/r)}_{\rm r}=\big(\bm{R}^{(b/r)}_{{\rm rx},b}\big)^{\frac{1}{2}}
  \bm{G}^{(b/r)}\big(\bm{R}^{(b/r)}_{{\rm tx},r}\big)^{\frac{1}{2}}$ is the scattered
 component of the Rician channel in which $\bm{R}^{(b/r)}_{{\rm rx},b}\in
 \mathbb{C}^{N\times N}$ is the spatial correlation matrix of the $N$ RAs at the BBU,
 $\bm{R}^{(b/r)}_{{\rm tx},r}\in \mathbb{C}^{K\times K}$ is the spatial correlation
 matrix of the $K$ forwarding TAs of the $R$ RRUs, and $\bm{G}^{(b/r)}\in
 \mathbb{C}^{N \times K}$ has i.i.d. complex entries and each of them has the distribution
 $\mathcal{CN}(0,1)$. The $r$th RRU has a total of $M$ antennas, which is much more than
 $U_r$, and it can always select its $U_r$ forwarding TAs to be spaced sufficiently far apart. Consequently,
 the correlations between the $U_r$ TAs can be assumed to be zero. Furthermore, there
 exists no correlation between the antennas of different RRUs. Thus we can assume that
 $\bm{R}^{(b/r)}_{{\rm tx},r}=\bm{I}_K$. 

 The uplink signals after combining can be expressed as
\begin{align}\label{eq34}
 \check{\bm{y}} =& \sqrt{\lambda^{(b/r)}} \bm{W}^{(b/r)} \bm{H}^{(b/r)}
  \big(\bm{P}_{\rm rx}^{(r,d)}\big)^{\frac{1}{2}} \bar{\bm{y}} + \check{\bm{n}} ,
\end{align}
 where $\bm{W}^{(b/r)}\in \mathbb{C}^{K\times N}$ is the BBU's combining matrix,
$\lambda^{(b/r)}=\frac{1}{K}\text{Tr}\Big\{\mathcal{E}\Big\{\bm{W}^{(b/r)}
 \big(\bm{W}^{(b/r)}\big)^{\rm H}\Big\}\Big\}$, and $\check{\bm{n}}=\sqrt{\lambda^{(b/r)}}
 \bm{W}^{(b/r)}\breve{\bm{n}}\in \mathbb{C}^K$ is the noise output after the
 combining, which obeys the distribution $\mathcal{CN}\big(\bm{0}_K,
 \bm{\Sigma}_{\check{n}}\big)$ with the covariance matrix $\bm{\Sigma}_{\check{n}}=
 \lambda^{(b/r)}\sigma_{\breve{n}}^{2}\bm{W}^{(b/r)}\left(\bm{W}^{(b/r)}\right)^{\rm H}$. 
 Again, the MF is adopted by the BBU and $\bm{W}^{(b/r)}$ is given by
\begin{align}\label{eq35}
 \bm{W}^{(b/r)} =& \left(\widehat{\bm{H}}^{(b/r)}\right)^{\rm H} ,
\end{align}
 where $\widehat{\bm{H}}^{(b/r)}$ is the estimate of the uplink channel $\bm{H}^{(b/r)}$.

 The MMSE channel estimate \cite{kay2003fundamentals} is given by (\ref{eq36}) at the top of
 this page, where $\breve{\bm{Y}}\in \mathbb{C}^{N\times K}$ is the received signal
 matrix with removing the LoS component over the $K$ pilot symbols before the receiver
 combining, and $\bm{X}^{(p)}_b\in
 \mathbb{C}^{K\times K}$ is the pilot symbol matrix with $\bm{X}^{(p)}_b
 \left(\bm{X}^{(p)}_b\right)^{\rm H}=\bm{I}_K$, while $\bm{P}_{{\rm rx}}^{(r, p)}=
 \text{diag}\big\{P_{{\rm rx},1}^{(r,p)},\cdots , P_{{\rm rx},K}^{(r,p)}\big\}$ are the
 received powers of the pilot symbols. In (\ref{eq36}), $\bm{\Psi}_{\bm{H}^{(b/r)}_{\rm r}}
 \in\mathbb{C}^{N K \times N K}$ is the covariance matrix of
 $\bm{vec}\left(\zeta\bm{H}^{(b/r)}_{\rm r}\right)$
 and it is given by\setcounter{equation}{36}
\begin{align}\label{eq37}
 \bm{\Psi}_{\bm{H}^{(b/r)}_{\rm r}} =& \mathcal{E}\left\{\bm{vec}\left(\zeta\bm{H}_{\rm r}^{(b/r)}\right)
  \bm{vec}\left(\zeta\bm{H}_{\rm r}^{(b/r)}\right)^{\rm H}\right\} \nonumber \\
 =& \zeta^2\bm{I}_K \otimes \bm{R}^{(b/r)}_{{\rm rx},b} .
\end{align}
 The distribution of the MMSE estimate $\bm{vec}\big(\widehat{\bm{H}}^{(b/r)}\big)$ is
 \cite{kay2003fundamentals}
\begin{align}\label{eq38}
 \bm{vec}\left(\widehat{\bm{H}}^{(b/r)}\right)\sim  \mathcal{CN}\left(\nu \bm{vec}\big(\bm{H}_{\rm d}^{(b/r)}\big),
  \bm{\Psi}_{\widehat{\bm{H}}^{(b/r)}}\right) ,
\end{align}
 with the covariance matrix formulated as
\begin{align}\label{eq39}
 \bm{\Psi}_{\widehat{\bm{H}}^{(b/r)}} =& \zeta^2\bm{\Psi}_{\bm{H}^{(b/r)}_{\rm r}} \Big(
  \sigma_{\breve{n}}^2 \big(\bm{P}_{{\rm rx}}^{(r,p)}\big)^{-1} \otimes \bm{I}_N +
  \zeta^2\bm{\Psi}_{\bm{H}^{(b/r)}_{\rm r}}\Big)^{-1} \nonumber \\
 & \times \zeta^2\bm{\Psi}_{\bm{H}^{(b/r)}_{\rm r}} .
\end{align}
 The relationship between the MMSE estimate $\bm{vec}\big(\widehat{\bm{H}}^{(b/r)}\big)$
 and the true channel $\bm{vec}\left(\bm{H}^{(b/r)}\right)$ is given by
\begin{align}\label{eq40}
 \bm{vec}\left(\bm{H}^{(b/r)}\right) =& \bm{vec}\left(\widehat{\bm{H}}^{(b/r)}\right)
  + \bm{vec}\left(\widetilde{\bm{H}}^{(b/r)}\right) , 
\end{align}
 where the channel estimation error $\bm{vec}\big(\widetilde{\bm{H}}^{(b/r)}\big)$ is
 statistically independent of both $\bm{vec}\big(\widehat{\bm{H}}^{(b/r)}\big)$ and
 $\bm{vec}\big(\bm{H}^{(b/r)}\big)$. Moreover, the distribution of
 $\bm{vec}\big(\widetilde{\bm{H}}^{(b/r)}\big)$ is given by
\begin{align}\label{eq41}
 \bm{vec}\big(\widetilde{\bm{H}}^{(b/r)}\big) \sim \mathcal{CN}\Big(\bm{0}_{N K},
  \bm{\Psi}_{\widetilde{\bm{H}}^{(b/r)}}\Big) ,
\end{align}
 with the covariance matrix formulated as
\begin{align}\label{eq42}
 \bm{\Psi}_{\widetilde{\bm{H}}^{(b/r)}} =& \bm{\Psi}_{\bm{H}^{(b/r)}_{\rm r}} -
  \bm{\Psi}_{\widehat{\bm{H}}^{(b/r)}} .
\end{align}

\begin{figure*}[bp!]\setcounter{equation}{54}
\vspace*{-3mm}
\hrulefill
\begin{align}\label{eq55}
 \bar{P}_{{\rm IN},k^*} =& a_{k^*} \lambda^{(b/r)} \lambda_{k^*}^{(r/u)} P_{{\rm rx},k^*}^{(r,d)}
  P_{{\rm rx},k^*}^{(u,d)} \Upsilon_{\rm IN,1} + \sum\limits_{k=1,j\neq k^*}^K
  a_{k^*} \lambda^{(b/r)} \lambda_{k}^{(r/u)} P_{{\rm rx},k}^{(r,d)} P_{{\rm rx},k}^{(u,d)}
  \Upsilon_{{\rm IN,2},k} \nonumber \\ &
  + \sum\limits_{k=1}^K \sum\limits_{j=1,j \neq k}^K \delta\big(\Delta_{j,k}\big)
  \lambda^{(b/r)} \lambda_{k}^{(r/u)} P_{{\rm rx},k}^{(r,d)} \Upsilon_{{\rm IN,3},k,j}
  + \Upsilon_{\rm IN,4} ,
\end{align}
\vspace*{-2mm}
\end{figure*}

\section{Achievable Throughput}\label{S3}

 The achievable throughput of a UD is determined by its final SINR at the BBU. There 
 are a total of $K$ UDs served by the BBU. Let us map the index of the $u_r$th UD served 
 by the $r$th RRU to $k$. That is, the $k$th UD, where $k \in \{1,2,\cdots,K\}$,
 is served by the $r$th RRU, and we can express $k$ as $k=\sum\nolimits_{r'=1}^{r-1}U_{r'}+u_r$.
 The SINR of the $k$th UD at the BBU is expressed as\setcounter{equation}{42}
\begin{align}\label{eq43}
 \gamma_k =& \frac{P_{{\rm S},k}}{P_{{\rm IN},k}} ,
\end{align}
 where $P_{{\rm S},k}$ is the power of the desired signal and $P_{{\rm IN},k}$ is the
 power of the interference plus noise. The ergodic achievable uplink throughput of the
 $k$th UD is defined as
\begin{align}\label{eq44}
 C_k = \mathcal{E}\left\{\log_2\left(1 + \gamma_{k}\right)\right\} .
\end{align}
 Recalling Lemma~\ref{L1} of Appendix~\ref{Apa}, the ergodic achievable uplink throughput 
 of the $k$th UD can be approximated as
\begin{align}\label{eq45}
 C_k \approx & \log_2\left(1 + \frac{\bar{P}_{{\rm S},k}}{\bar{P}_{{\rm IN},k}}\right) ,
\end{align}
 where $\bar{P}_{{\rm S},k}=\mathcal{E}\left\{P_{{\rm S},k}\right\}$ and
 $\bar{P}_{{\rm IN},k}=\mathcal{E}\left\{P_{{\rm IN},k}\right\}$.

 To calculate the ergodic signal power $\bar{P}_{{\rm S},k}$ and the interference plus
 noise power $\bar{P}_{{\rm IN},k}$, we substitute (\ref{eq21}) into (\ref{eq34}) to
 arrive at the $k^{*}$th UD's signal after the BBU's combining operation as given in
 (\ref{eq46}) at the bottom of the previous page, where $a_{k^*}$,
 $P_{{\rm rx},k^*}^{(u,d)}$ and $P_{{\rm rx},k^*}^{(r,d)}$ are the $k^*$th diagonal
 elements of $\bm{A}$, $\bm{P}_{\rm rx}^{(u,d)}$ and $\bm{P}_{\rm rx}^{(r,d)}$,
 respectively. Furthermore, $\widetilde{s}_{r',j}$ is the $j$th element of
 $\widetilde{\bm{s}}_{r'}$, $\check{\bar{n}}_{k^*}=\sqrt{a_{k^*}\lambda_{k^*}^{(r/u)}
 P_{{\rm rx},k^*}^{(r,d)}} \bm{W}_{[k^*:\,]}^{(b/r)}\bm{H}_{[\,:k^*]}^{(b/r)} n_{k^*}
 +\check{n}_{k^*}$ is the effective noise, and the $k^*$th UD is served by the $r^*$th
 RRU. Moreover, $\delta\big(\Delta_{j,k}\big)$ in (\ref{eq46}) is the Dirac delta
 function, and the  value of $\Delta_{j,k}$ depends on whether the $j$th and $k$th
 UDs are served by the same RRU. Specifically, if both the $j$th and $k$th UDs are
 served by the same RRU, we have $\Delta_{j,k}=1$; otherwise $\Delta_{j,k}=0$.

 Then,  the signal power $\bar{P}_{{\rm S},k^{*}}$ is given in (\ref{eq47}) at the bottom of
 the previous page. Intuitively, the UD-to-RRU channel links are independent of the RRU-to-BBU
 channel links. Hence, we can calculate their expectation separately.
 By denoting \setcounter{equation}{47}
\begin{align}
 \Upsilon_{s,1} =& \mathcal{E}\left\{\Big(\widehat{\bm{H}}_{[\,:k^*]}^{(r/u)}\Big)^{\rm H}
  \bm{H}_{[\,:k^*]}^{(r/u)}\Big(\bm{H}_{[\,:k^*]}^{(r/u)}\Big)^{\rm H}
  \widehat{\bm{H}}_{[\,:k^*]}^{(r/u)}\right\} , \label{eq48} 
\end{align}
\begin{align}
 \Upsilon_{s,2} =& \mathcal{E}\left\{\Big(\widehat{\bm{H}}_{[\,:k^*]}^{(b/r)}\Big)^{\rm H}
  \widehat{\bm{H}}_{[\,:k^*]}^{(b/r)}\Big(\widehat{\bm{H}}_{[\,:k^*]}^{(b/r)}\Big)^{\rm H}
  \widehat{\bm{H}}_{[k^*:\,]}^{(b/r)}\right\} , \label{eq49}
\end{align}
 $\bar{P}_{\text{S},k^*}$ can be expressed as
\begin{align}\label{eq50}
 \bar{P}_{\text{S},k^{*}} =& a_{k^*} \lambda^{(b/r)} \lambda_{k^*}^{(r/u)}
  P_{\text{rx},k^{*}}^{(r,d)} P_{\text{rx},k^{*}}^{(u,d)} \Upsilon_{s,1} \Upsilon_{s,2} .
\end{align}
 Recalling Lemmas~\ref{L2} and \ref{L3} of Appendix~\ref{Apa}, we have
\begin{align}\label{eq51}
 \Upsilon_{s,1} =& \mathcal{E}\Big\{ \Big(\widehat{\bm{H}}_{[\,:k^{*}]}^{(r/u)}\Big)^{\rm H}
  \Big(\widehat{\bm{H}}_{[\,:k^{*}]}^{(r/u)} + \widetilde{\bm{H}}^{(r/u)}_{[\,:k^{*}]}\Big) \nonumber \\
 & \times \Big(\widehat{\bm{H}}_{[\,:k^{*}]}^{(r/u)} + \widetilde{\bm{H}}_{[\,:k^{*}]}^{(r/u)}\Big)^{\rm H}
  \widehat{\bm{H}}_{[\,:k^{*}]}^{(r/u)}\Big\} \nonumber \\
 =& \Big(\text{Tr}\Big\{\bm{\Psi}_{\widehat{\bm{H}}^{(r/u)}_{[\,:k^{*}]}}\Big\}\Big)^2 +
  \text{Tr}\Big\{\bm{\Psi}_{\widehat{\bm{H}}^{(r/u)}_{[\,:k^{*}]}} 
  \bm{\Psi}_{\widetilde{\bm{H}}^{(r/u)}_{[\,:k^{*}]}}\Big\} ,
\end{align}
 where $\bm{\Psi}_{\widehat{\bm{H}}^{(r/u)}_{[\,:k^{*}]}}$ and
 $ \bm{\Psi}_{\widetilde{\bm{H}}^{(r/u)}_{[\,:k^{*}]}}$ denote the covariance matrices
 of $\widehat{\bm{H}}^{(r/u)}_{[\,:k^{*}]}$ and $\widetilde{\bm{H}}^{(r/u)}_{[\,:k^{*}]}$,
 respectively. Recalling Lemma~\ref{L3} of Appendix~\ref{Apa}, we have
\begin{align}\label{eq52}
 \Upsilon_{s,2} =& \Big(\text{Tr}\Big\{\nu^2 \bm{B}_{\bm{H}_{{\rm d} ~ [\,:,k^{*}]}^{(b/r)}} +
   \bm{\Psi}_{\widehat{\bm{H}}_{{\rm r} ~ [\,:k^{*}]}^{(b/r)}}\Big\}\Big)^2 ,
\end{align}
 where we have $\bm{B}_{\bm{H}_{{\rm d} ~ [\,:,k^{*}]}^{(b/r)}}\! =\! \bm{H}_{{\rm d} ~ [\,:,k^{*}]}^{(b/r)}
 \Big(\bm{H}_{{\rm d} ~ [\,:,k^{*}]}^{(b/r)}\Big)^{\rm H}$ and
 $\bm{\Psi}_{\widehat{\bm{H}}_{{\rm r} ~ [\,:k^{*}]}^{(b/r)}}$ is the covariance
  matrix of $\widehat{\bm{H}}_{{\rm r} ~ [\,:k^{*}]}^{(b/r)}$. Furthermore, $\lambda^{(b/r)}$
 and $\lambda_{k^{*}}^{(r/u)}$ are given respectively by
\begin{align}
 \lambda^{(b/r)} =& \left(\frac{1}{K}\text{Tr}\Big\{\nu^2 \bm{B}_{\bm{H}_{{\rm d}}^{(b/r)}}
  +  \bm{\Psi}_{\bm{H}_{{\rm r}}^{(b/r)}}\Big\}\right)^{-1} , \label{eq53} \\
 \lambda_{k^*}^{(r/u)} =& \left(\frac{1}{U_{r^*}}\text{Tr}\Big\{\bm{\Psi}_{\bm{H}_{{r}}^{(r/u)}}
  \Big\}\right)^{-1} , \label{eq54}
\end{align}
 where $\bm{B}_{\bm{H}_{{\rm d}}^{(b/r)}}=\bm{vec}\big(\bm{H}_{{\rm d}}^{(b/r)}\big)
 \bm{vec}\big(\bm{H}_{{\rm d}}^{(b/r)}\big)^{\rm H}$. By substituting (\ref{eq51}) to
 (\ref{eq54}) into (\ref{eq50}), we can calculate $\bar{P}_{\text{S},k^{*}}$.

 The interference plus noise power $\bar{P}_{{\rm IN},k}$ is given in (\ref{eq55}) at the 
 bottom of this page, in which $\Upsilon_{\rm IN,1}$, $\Upsilon_{{\rm IN,2},k}$,
 $\Upsilon_{{\rm IN,3},k,j}$ and $\Upsilon_{\rm IN,4}$ are defined respectively by\setcounter{equation}{55}
\begin{align}
 \Upsilon_{\rm IN,1} =& \mathcal{E}\Big\{ \Big| \bm{W}_{[k^*:\,]}^{(b/r)} \widetilde{\bm{H}}_{[\,:k^*]}^{(b/r)}
  \bm{W}_{[k^*:\,]}^{(r/u)} \bm{H}_{[\,:k^*]}^{(r/u)}\Big|^2 \Big\} , \label{eq56} \\
 \Upsilon_{{\rm IN,2},k} =& \mathcal{E}\Big\{\Big| \bm{W}_{[k^*:\,]}^{(b/r)}\bm{H}_{[\,:k]}^{(b/r)}
  \bm{W}_{[k:\,]}^{(r/u)} \bm{H}_{[\,:k]}^{(r/u)}\Big|^{2}\Big\} , \label{eq57} \\
 \Upsilon_{{\rm IN,3},k,j} =& \mathcal{E}\bigg\{\bigg| \bm{W}_{[k^*:\,]}^{(b/r)}
  \bm{H}_{[\,:k]}^{(b/r)} \bm{W}_{[k:\,]}^{(r/u)} \sum\limits_{r'=2}^R\bm{V}^{(r/u)}_{r' ~ [\,:j]}\bigg|^2\bigg\}
  , \label{eq58} \\
 \Upsilon_{\rm IN,4} =& \mathcal{E}\Big\{ \Big|\check{\bar{n}}_{k^*}\Big|^{2} \Big\} . \label{eq59}
\end{align}
 Recalling Lemma~\ref{L2} of Appendix~\ref{Apa}, we have $\Upsilon_{\rm IN,1}$ given by
 (\ref{eq60}). Similarly, recalling Lemmas~\ref{L2} and \ref{L3}, we can express
 $\Upsilon_{{\rm IN,2},k}$ as given in (\ref{eq61}). Furthermore, as shown in Appendix~\ref{Apb},
 $\Upsilon_{{\rm IN,3},k,j}$ is given in (\ref{eq62}), in which $\bm{\Psi}_{\Sigma\bm{V}^{(r/u)}_{r'\,[\,:j]}}$
 is expressed in (\ref{be4}) of Appendix~\ref{Apb}, while $\Upsilon_{\rm IN,4}$ can 
 be calculated according to (\ref{eq63}). The equations (\ref{eq60}) to (\ref{eq63}) are all
 listed at the top of the next page. By substituting (\ref{eq60}) to (\ref{eq63}) into (\ref{eq55}),
 we can calculate $\bar{P}_{\text{IN},k^*}$.

 Finally, substituting (\ref{eq50}) and (\ref{eq55}) into (\ref{eq45}) leads to
 the ergodic achievable uplink throughput $C_{k^*}$.

\begin{figure*}[tp!]\setcounter{equation}{59}
\vspace*{-2mm}
\begin{align}
 \Upsilon_{\rm IN,1}=& \bigg( \Big(\text{Tr}\Big\{\bm{\Psi}_{\widehat{\bm{H}}^{(r/u)}_{[\,:k^*]}}\Big\}
  \Big)^2 + \text{Tr}\Big\{\bm{\Psi}_{\widehat{\bm{H}}^{(r/u)}_{[\,:k^*]}}
  \bm{\Psi}_{\widetilde{\bm{H}}^{(r/u)}_{[\,:k^*]}}\Big\} \bigg)
  \text{Tr}\Big\{\bm{\Psi}_{\widetilde{\bm{H}}_{[\,:k^*]}^{(b/r)}} \Big(\nu^2
  \bm{B}_{\bm{H}_{{\rm d} ~ [\,:k^*]}^{(b/r)}} + 
  \bm{\Psi}_{\widehat{\bm{H}}_{{\rm r} ~ [\,:k^*]}^{(b/r)}}\Big)\Big\} . \label{eq60} \\
 \Upsilon_{{\rm IN,2},k}  =& \bigg( \Big(\text{Tr}\Big\{\bm{\Psi}_{\widehat{\bm{H}}^{(r/u)}_{[\,:k]}}
  \Big\}\Big)^2 +\text{Tr}\Big\{\bm{\Psi}_{\widehat{\bm{H}}^{(r/u)}_{[\,:k]}}
  \bm{\Psi}_{\widetilde{\bm{H}}^{(r/u)}_{[\,:k]}}\Big\} \bigg) \nonumber \\
 & \times \bigg( \text{Tr}\bigg\{ \Big(\nu^2 \bm{B}_{\bm{H}_{{\rm d} ~ [\,:k^*]}^{(b/r)}} + 
  \bm{\Psi}_{\widehat{\bm{H}}_{{\rm r} ~ [\,:k^*]}^{(b/r)}}\Big) \Big(\nu^2
  \bm{B}_{\bm{H}_{{\rm d} ~ [\,:k]}^{(b/r)}} +  \bm{\Psi}_{\bm{H}_{{\rm r} ~ [\,:k]}^{(b/r)}}\Big)
  \bigg\} \bigg) . \label{eq61} \\
 \Upsilon_{{\rm IN,3},k,j} =& \text{Tr}\bigg\{ \bm{\Psi}_{ \widehat{\bm{H}}^{(r/u)}_{[\,:k]} }
  \bm{\Psi}_{ \Sigma\bm{V}^{(r/u)}_{r'\,[\,:j]} } \bigg\} \text{Tr}\bigg\{ \bigg( \nu^2
  \bm{B}_{ \bm{H}_{{\rm d} ~ [\,:k^*]} }^{(b/r)} +  \bm{\Psi}_{ \widehat{\bm{H}}_{{\rm r} ~[\,:k^*]} }^{(b/r)}
  \bigg) \bigg( \nu^2 \bm{B}_{ \bm{H}_{{\rm d} ~ [\,:k]} }^{(b/r)} + 
  \bm{\Psi}_{ \bm{H}_{{\rm r} ~ [\,:k]} }^{(b/r)} \bigg) \bigg\} . \label{eq62} \\
\label{eq63}
 \Upsilon_{\rm IN,4} =& a_{k^*} \lambda^{(b/r)} \lambda_{k^*}^{(r/u)} P_{\text{rx},k^*}^{(r,d)}
  \sigma_{\widetilde{n}}^2 \text{Tr}\bigg\{\bm{\Psi}_{\widehat{\bm{H}}^{(r/u)}_{[\,:k^*]}}\bigg\}
  \text{Tr}\bigg\{\bigg(\nu^2 \bm{B}_{\bm{H}_{{\rm d} ~ [\,:,k^*]}^{(b/r)}} + 
  \bm{\Psi}_{{\bm{H}}_{{\rm r} ~ [\,:k^*]}^{(b/r)}}\bigg) \bigg(\nu^2 \bm{B}_{\bm{H}_{{\rm d} ~ [\,:,k^*]}^{(b/r)}}
  +  \bm{\Psi}_{\widehat{\bm{H}}_{{\rm r} ~ [\,:k^*]}^{(b/r)}}\bigg)\bigg\} \nonumber \\
 & + \lambda^{(b/r)} \text{Tr}\bigg\{\nu^2 \bm{B}_{\bm{H}_{{\rm d} ~ [\,:,k^*]}^{(b/r)}} + 
   \bm{\Psi}_{\widehat{\bm{H}}_{{\rm r} ~ [\,:k^*]}^{(b/r)}}\bigg\} \sigma_{\breve{n}}^2 .
\end{align}
\hrulefill
\vspace*{-5mm}
\end{figure*}

\section{Optimal Power sharing for UDs and RRUs}\label{S4}

 The achievable uplink throughput depends on the accuracy of channel estimate, which
 in turn is dominated by the power allocated to pilots. Again, more power allocated
 to pilots will result in more accurate channel estimates, which will enhance
 the achievable throughput. However, the achievable uplink throughput also depends on
 the power allocated to data transmission. Allocating more power to pilots 
 will in turn reduce the power allocated to data transmission, which will reduce the
 achievable throughput. Therefore, intuitively, for each UD, there exists an `optimal'
 power sharing between its pilots and data transmission. Likewise, for every
 RRU, there exists an `optimal' power sharing between its forwarding-TAs' pilots
   and data forwarding transmission. Hence, in order to maximize the system's
 uplink throughput, we have to jointly optimize the power sharing between pilot
 training and data transmission for all the UDs and their host RRUs.

 Mathematically, for the $k^*$th UD, its achievable uplink throughput $C_{k^*}$ is a
 function of the power sharing factors for all the UDs and all the RRUs in the
 system, that is, $C_{k^*}=C_{k^*}(\bm{\eta})$, where\setcounter{equation}{63}
\begin{align}\label{eq64}
 \bm{\eta} =& \Big[\eta_1^{(r/u)} ~ \eta_2^{(r/u)} \cdots \eta_K^{(r/u)} ~ \eta_1^{(b/r)} ~
  \eta_2^{(b/r)} \cdots \eta_K^{(b/r)}\Big]^{\rm T} .
\end{align}
 The sum-rate of all the $K$ UDs served by the host BBU, which is given by
\begin{align}\label{eq65}
 C_{\text{sum}}(\bm{\eta}) = & \sum\limits_{k^*=1}^K C_{k^*}(\bm{\eta}) ,
\end{align}
 is a function of the power sharing factor set $\bm{\eta}$ for all the UDs and all
 the RRUs. In order to boost the achievable sum-rate, we can optimize the power
 sharing factor set $\bm{\eta}$ by maximizing $C_{\text{sum}}$ subject to the
 constraints of $0 < \eta_k^{(r/u)} < 1$ and $0 < \eta_k^{(b/r)} < 1$ for $1\le k\le K$. 
 Thus, we can formulate the global optimization of the power sharing for the
 C-RAN uplink as the following optimization problem
\begin{align}\label{eq66}
\begin{array}{cl}
 \bm{\eta}_{\text{opt}} = & \max\limits_{\bm{\eta}}\, C_{\text{sum}}(\bm{\eta}) ,  \\
 \text{s.t.}: &  0 < \eta_k^{(r/u)} < 1 , ~ 1\le k\le K ,  \\
              &  0 < \eta_k^{(b/r)} < 1 , ~ 1\le k\le K . 
\end{array}
\end{align}

The above problem is a multivariate optimization problem, where the cost function
 $C_{\text{sum}}(\bm{\eta})$ is the summation of $K$ log-functions having $2K$
 multivariate factors.  Furthermore, the underlying system is a two-layer network,
 consisting the multiple UDs-to-RRUs links at the first layer and the multiple
 RRUs-to-BBC links at the second layer. In particular, as observed in (\ref{eq46}), there
 is residual inter-UD interference between the $K$ UDs. Consequently, the $K$ decision
 variables $\eta_k^{(b/r)}$ for $1\le k\le K$ are the functions of the other $K$
 decision variables $\eta_k^{(r/u)}$ for $1\le k\le K$. Hence, the $2K$ optimization
 decision variables are dependent on each other. It is therefore impossible to
 derive a closed-form solution to this complex multivariate optimization problem, and 
 a numerical solution must be sought. More specifically, the Karush-Kuhn-Tucker (KKT)
 conditions associated with the constrained optimization problem (\ref{eq66}) are highly 
 complex, from which no closed-form solution can be derived.  Additionally, at the time of writing it remains an open question, whether (\ref{eq66}) is convex or not.

\begin{algorithm}
\caption{Differential evolutionary algorithm}
\label{Alg1}
\begin{algorithmic}[1]
\State Set generation index $g = 0$ and randomly generate the initial population
 of $P_{s}$ individuals $\bm{\eta}^{(g,p_{s})}, p_{s} \in \{1,2,\cdots,P_{s}\}$.
 \Comment{Initialization.}
\While{$g < G_{\text{max}}$}          
  \For{$p_{s} = 1$ to $P_{s}$} \Comment{Mutation}
    \State Create a mutated vector $\tilde{\bm{\eta}}^{(g,p_{s})} = \bm{\eta}^{(g,p_{s})}
      + \lambda_{p_{s}}(\check{\bm{\eta}}^{(g,p_{s})} - \bm{\eta}^{(g,p_{s})}) +
      \lambda_{p_{s}}(\bm{\eta}^{(g,p'_{s})} - \bm{\eta}^{(g,p''_{s})})$.
   \EndFor
   \For{$p_{s} = 1$ to $P_{s}$} \Comment{Crossover}
     \For{$\alpha = 1$ to $A$}
       \State Crossover the $\alpha$th element of $\bm{\eta}^{(g,p_{s})}$ by 
         \begin{align}
	   \breve{\eta}^{(g,p_{s})}_{\alpha} =& \left\{\begin{array}{ll} \tilde{\eta}^{(g,p_{s})}_{\alpha},
             & \text{rand}_{\alpha}(0,1) \le C_{r_{p_s}} , \\
	   \eta^{(g,p_{s})}_{\alpha}, & \text{otherwise}, \end{array}\right. \nonumber
         \end{align}
     \EndFor
   \EndFor
   \For{$p = 1$ to $P_{s}$} \Comment{Selection}
     \State Select $\bm{\eta}^{(g,p_{s})}$ or $\breve{\bm{\eta}}^{(g,p_{s})}$ surviving to the next generation
     \begin{align}
       \bm{\eta}^{(g + 1,p_{s})} =& \left\{ \begin{array}{cl} \breve{\bm{\eta}}^{(g,p_{s})},
       & C_{\text{sum}}(\bm{\eta}^{(g,p_{s})}) \le C_{\text{sum}}(\breve{\bm{\eta}}^{(g,p_{s})}) , \\
       \bm{\eta}^{(g,p_{s})}, & \text{otherwise} .\end{array}\right.\!\!\! \nonumber
     \end{align}
  \EndFor      
\EndWhile
\end{algorithmic}
\end{algorithm}

 Solving the optimization problem (\ref{eq66}) using numerical
 optimization algorithms, such as the expectation maximization algorithm \cite{zhang2011joint},
 the repeated weighted boosting search algorithm \cite{chen2005experiments,zhang2011jointchannel,RWB2012} and evolutionary
 algorithms \cite{evolutionary2013Conf,zhang2014evolutionary,DEA2018}, will impose considerable computational complexity. Fortunately,
 in the C-RAN architecture, the BBUs cooperate across the BBU pool, which is an
 aggregated collective resource shared among a large number of virtual BSs. As a benefit, the system becomes 
 capable of achieving a much improved exploitation of the processing resources, while
 imposing a reduced power consumption, based on statistical computing multiplexing \cite{pompili2015dynamic}.
 Therefore, the C-RAN possesses a powerful centralized signal processing
 capability for dynamic shared resource allocation \cite{pan2017joint-twc}. In
 other words, the C-RAN has the necessary computing power for numerically solving the optimization
 problem (\ref{eq66}).

  Since it is unknown, whether the optimization problem (\ref{eq66})
 is convex or not, we invoke the continuous-variable DEA  \cite{price2005differential,qin2009differential} 
 for finding the globally optimal solution $\bm{\eta}_{\text{opt}}$. The pseudo-code
 of the DEA is presented in Algorithm~1, where $P_s$ is the population size,
 $G_{\text{max}}$ is the pre-set maximum number of generations, $\tilde{\bm{\eta}}^{(g,p_s)}$
 represents a mutated vector, $\breve{\bm{\eta}}^{(g,p_s)}$ is a trial vector,
 $\check{\bm{\eta}}^{(g,p_{s})}$ is selected from an elite-archive having high-fitness individuals and finally 
 $\lambda_{p_s}\in (0, ~ 1]$ is a randomly generated scaling factor. Furthermore, $A$ is the total number of 
 elements in vector $\bm{\eta}^{(g,p_s)}$, $\text{rand}_{\alpha}(0,1)$ denotes the
 random number drawn  from the uniform distribution in $[0, ~ 1]$ for the $\alpha$-th
 element, while $C_{r_{p_s}}\in [0,~ 1]$ is the crossover probability. Additionally, 
 the symbol $\rhd$ denotes a comment. The detailed characteristics of this
 DEA can be found in \cite{evolutionary2013Conf,zhang2014evolutionary,DEA2018}.

\begin{table}[bp!]
\vspace*{-5mm}
\caption{Default system parameters of the massive MIMO aided C-RAN.}
\vspace*{-4mm}
\begin{center}
\begin{tabular}{l|l}
\hline\hline
 Bandwidth & 10 MHz \\ \hline
 Number of UDs $K$ & 10 \\ \hline
 Number of RRUs $R$ & 4 \\ \hline
 Number of RAs at BBU $N$ & 128 \\ \hline
 Number of RAs at RRU $M$ & 32 \\ \hline
 Spatial correlation factor of RAs $\rho$ & 0.1 \\ \hline
 Rician factor $K_{Rice}$ & 10 dB \\ \hline
 Receiver efficiency factor $\rho_r$ & 0.1 \\ \hline
 Noise power spectral density & -174 dBm/Hz \\ \hline
 Total power of a single UD $P_{\rm UD}$ & 0.2 Watts \\ \hline
 Total power of a single RRU $P_{\rm RRU}$ & 10 Watts \\ \hline\hline
\end{tabular}
\end{center}
\label{Tab1}
\vspace*{-2mm}
\end{table}

\section{Simulation Study}\label{S5}

 In the simulated massive MIMO aided C-RAN, the coverage of the BBU is a square area of 
 $[-500\sqrt{2}, ~ 500\sqrt{2}] \times [-500\sqrt{2}, ~ 500\sqrt{2}]$ square meters
 (${\rm m}^2$), and the BBU is located at the center of the square. $R=4$ RRUs are
 deployed at the coordinates of $(250\sqrt{2},250\sqrt{2})$\,m, $(250\sqrt{2},
 -250\sqrt{2})$\,m, $(-250\sqrt{2},-250\sqrt{2})$\,m, and $(-250\sqrt{2},250\sqrt{2})$\,m,
 respectively, by default. Thus, the distance between a RRU and the BBU is 500\,m by
 default. $K$ UDs are independently and uniformly distributed in this square area. Each
 UD is equipped with a single antenna, and each RRU is equipped with $M = 32$ RAs, whilst
 the BBU employs $N=128$ RAs.  Naturally, the number of TAs at each RRU
 is an integer. For reasons of fairness, $U_r$ for $1\le r\le R$ should be similar and they are all close to $K/R$ subject to the constraint $\sum_{r=1}^RU_r=K$. Consider
 for example the default system associated with a total number of  $K=10$ UDs, which are served
 by $R=4$ RRUs. A possible choice is $(U_1,U_2,U_3,U_4)=(2,2,3,3)$, which is used in
 the  simulations. The $m$th-row and $n$th-column element of the spatial
 correlation matrix $\bm{R}_{{\rm rx},r}^{(r/u)}$ of the $r$th RRU's RAs is given by
 $\rho^{|m-n|}$, where $\rho$ is the spatial correlation  factor of the RAs. Similarly,
 the $m$th-row and $n$th-column element of the spatial correlation matrix
 $\bm{R}_{{\rm rx},b}^{(b/r)}$ of the BBU's RAs is also specified by $\rho^{|m-n|}$. The
 system's bandwidth is 10\,MHz, and the noise power spectral density is -174\,dBm/Hz. The
 power allocated to a single UD is $P_{\rm UD}=0.2$\,Watts, and the power allocated to a
 single RRU is $P_{\rm RRU}=10$\,Watts. The default system parameters of this simulated
 massive MIMO aided C-RAN are listed in Table~\ref{Tab1}. Without optimization, the power
 sharing factors $\eta_k^{(r/u)}$ and $\eta_k^{(b/r)}$ are all set to 0.5, for $k\in\{1,2,\cdots ,K\}$,
 i.e., half power for pilot training and half for data transmission/forwarding.

  It is well-known that as an efficient global optimization algorithm, the DEA is
 capable of converging fast to an optimal solution of a complex optimization problem. The 
 efficiency, reliability, and convergence characteristics of the DEA have been extensively
 investigated and documented in \cite{evolutionary2013Conf,zhang2014evolutionary,DEA2018}.
 Hence, we focus on investigating the influence of the system's key parameters on the
 achievable sum-rate. These key parameters are the number of RAs at a RRU $M$, the number
 of RAs at the BBU $N$, the number of UDs $K$ being served simultaneously, the number of
 RRUs $R$, the correlation factor of RAs $\rho$, and the Rician factor $K_{Rice}$. In the
 following, `Without power optimization' indicates the sum-rate  calculated using
 (\ref{eq65}) with the non-optimized power sharing factors for UDs and RRUs, i.e.,
 $\eta_k^{(r/u)}=0.5$ and $\eta_k^{(b/r)}=0.5$ for $1\le k\le K$, whilst `With power
 optimization' is the sum-rate using (\ref{eq65}) with  the globally optimized power
 sharing factors obtained by solving the optimization problem  (\ref{eq66}) using the
 DEA. Furthermore, `Theoretical' indicates the sum-rate calculated using (\ref{eq65}),
 while `Simulation' is the Monte-Carlo simulation result obtained by averaging over 400
 random realizations.

\begin{figure}[tp!]
\begin{center}
 \includegraphics[width=0.48\textwidth,angle=0]{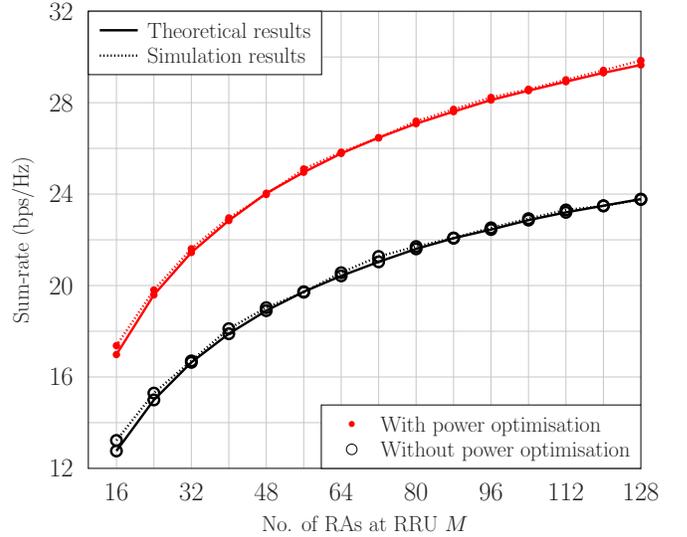}
\end{center}
\vspace{-5mm}
\caption{The sum-rate as a function of the number of RRU's RAs $M$. The rest of the
 system parameters are as given in Table~\ref{Tab1}.}
\label{FIG2}
\vspace{-5mm}
\end{figure}

 Fig.~\ref{FIG2} depicts the achievable sum-rate as the function of the number of
 RAs $M$ deployed at each RRU. It is widely recognized that for `single-layer' massive
 MIMO systems associated with Rayleigh channel matrices, the sum rate is a smoothly increasing
 function of the number of receiver antennas. This smooth shape obeys the logarithmic function of the system's SINR improvement upon increasing receiver antennas. Our massive MIMO aided C-RAN is
 a `two-layer' network with the first layer being Rayleigh and the second layer being
 Rician distributed. In the scenario of Fig.~2, the second layer is fixed, hence we can express the
 sum rate of (\ref{eq65}) equivalently as
\begin{align}\label{eq67}
 C(M) =& \sum_{k=1}^K \log_2\big(1 + \text{SINR}_k(M) \big) ,
\end{align}
 where $\text{SINR}_k(M)$ is the $k$th UD's SINR for this network, while $M$ is the number
 of RAs at each RRU. Changing $M$ changes the Rayleigh channel matrices of the first
 layer. Since  this `equivalent' system is Rayleigh, $C(M)$ is an increasing function
 of $M$ having a smooth shape determined by the logarithmic function of the system's
 SINR. This is confirmed by the results of Fig.~\ref{FIG2}.  It can also be seen from
 Fig.~2 that the achievable sum-rate associated with power sharing optimization is much
 higher than that without power sharing optimization. Specifically, the sum-rate gain
 achieved by the proposed globally optimal power sharing is about 33\%, which increases from
 4.2\,bps/Hz to 5.9\,bps/Hz, when $M$ is increased from 16 to 128. 

\begin{figure}[tp!]
\vspace{0mm}
\begin{center}
 \includegraphics[width=0.48\textwidth,angle=0]{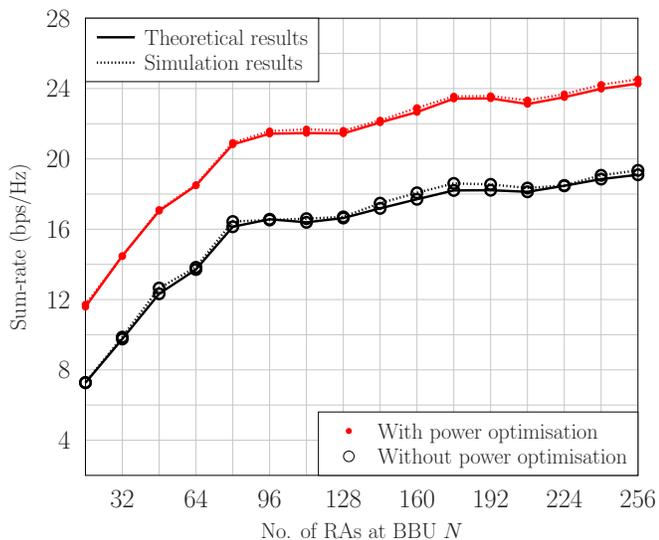}
\end{center}
\vspace{-5mm}
\caption{The sum-rate as a function of the number of BBU's RAs $N$. The rest of the
 system parameters are as given in Table~\ref{Tab1}.}
\label{FIG3}
\vspace{-5mm}
\end{figure}

Fig.~\ref{FIG3} portrays the achievable sum-rate as a function of the number of RAs $N$
 deployed at the BBU. It can be seen that the massive MIMO aided C-RAN associated
 with the proposed global optimization of power sharing is capable of offering a
 sum-rate gain from 4.3\,bps/Hz to 5.2\,bps/Hz, when $N$ increases from
 16 to 256, compared to the  same system operating without power sharing optimization.
  Observe from Fig.~\ref{FIG3} that the sum rate is an increasing function of $N$, but
 its shape is not very smooth. This phenomenon may be attributed to the LoS
 path of the Rician channel. In the scenario of Fig.~\ref{FIG3}, the first layer is fixed,
 and the sum rate of (\ref{eq65}) can be equivalently expressed as
\begin{align}\label{eq68}
 C(N) =& \sum_{k=1}^K \log_2\big(1 + \text{SINR}_k(N) \big),
\end{align}
 where $\text{SINR}_k(N)$ represents the $k$th UD's SINR for this network, while $N$ is
 the number of RAs at the BBC. Note that in contrast to the scenario of Fig.~\ref{FIG2}, this 
 `equivalent' system is Rician. Our experience with `single-layer' massive MIMO scenarios associated with
 Rician channel matrices \cite{zhang2018adaptive,zhang2018regularized} shows that the sum rate is an increasing function of
 the number of RAs but it exhibits slight undulations, similar
 to those seen in Fig.~\ref{FIG3}.

\begin{figure}[tp!]
\vspace{-0mm}
\begin{center}
 \includegraphics[width=0.48\textwidth,angle=0]{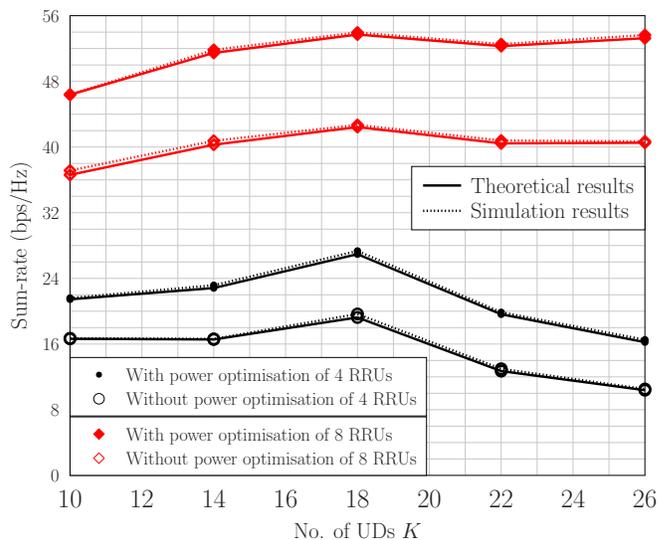}
\end{center}
\vspace{-5mm}
\caption{The sum-rates of the 4-RRU and 8-RRU systems as a functions of the number of
 UDs $K$. The rest of the system parameters are as given in Table~\ref{Tab1}.}
\label{FIG4}
\vspace{-5mm}
\end{figure}

\begin{figure}[tp!]
\vspace{0mm}
\begin{center}
 \includegraphics[width=0.48\textwidth,angle=0]{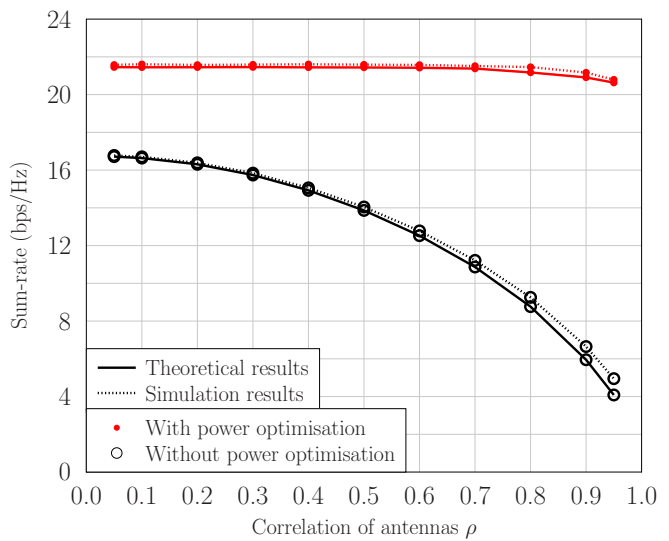}
\end{center}
\vspace{-5mm}
\caption{The sum-rate as a function of the spatial correlation factor $\rho$ of RAs. The
 rest of the system parameters are as given in Table~\ref{Tab1}.}
\label{FIG5}
\vspace{-5mm}
\end{figure}

 The joint impact of both the number of RRUs and the number of UDs is investigated
 next. We consider the two systems, the default system having $R=4$ RRUs and the
 system associated with $R=8$ RRUs. In the latter case, we divide the coverage area
 of the BBU into 8 equal-size subareas and place a RRU at the center of each subarea.
 Fig.~\ref{FIG4} compares the achievable sum-rates of the 4-RRU and 8-RRU systems as
 the functions of $K$. As expected, the sum-rate of the 8-RRU system is considerably
 higher than that of the 4-RRU system. Furthermore, higher sum-rate gain is attained
 by the proposed global optimization for the 8-RRU system, than for the 4-RRU system
 Additionally, it is seen from Fig.~\ref{FIG4} that the sum-rate increases as the
 number of UDs increases for small $K$ but for large $K$, the sum-rate becomes
 decreasing as as $K$ increases further. This is because for large $K$, user
 interference becomes dominant, which reduces the achievable throughput. 

\begin{figure}[tp!]
\vspace{0mm}
\begin{center}
 \includegraphics[width=0.48\textwidth,angle=0]{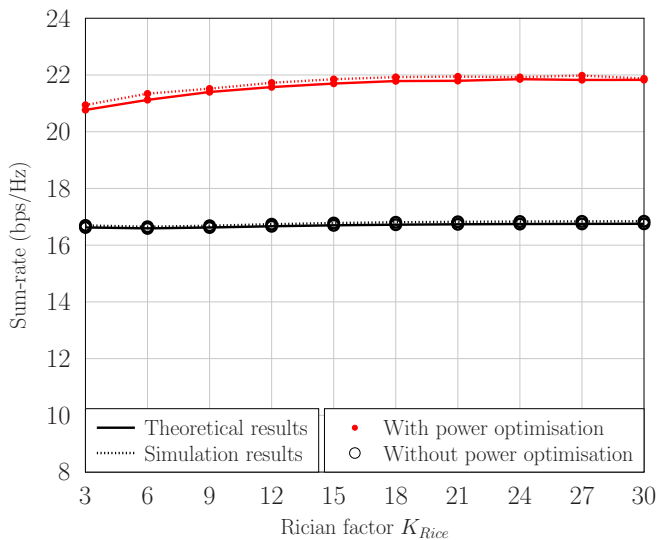}
\end{center}
\vspace{-5mm}
\caption{The sum-rate as a function of the Rician factor $K_{Rice}$. The rest of the
 system parameters are as given in Table~\ref{Tab1}.}
\label{FIG6}
\vspace{-5mm}
\end{figure}

\begin{figure*}[bp!]\setcounter{equation}{72}
\vspace*{-4mm}
\hrulefill
\begin{align}\label{be1} 
 \Upsilon_{{\rm IN,3},k,j}   =& \text{Tr}\left\{ \mathcal{E}\left\{ \sum\limits_{r'=2}^R \bm{V}^{(r/u)}_{r' ~ [\,:j]}
  \bigg(\sum\limits_{r'=2}^R \bm{V}^{(r/u)}_{r' ~ [\,:j]}\bigg)^{\rm H} \Big(\bm{W}_{[k:\,]}^{(r/u)}\Big)^{\rm H}
  \bm{W}_{[k:\,]}^{(r/u)}\right\}\right\} \text{Tr}\left\{ \mathcal{E}\bigg\{ \Big(\bm{W}_{[k^*:\,]}^{(b/r)}\Big)^{\rm H}
  \bm{W}_{[k^*:\,]}^{(b/r)} \bm{H}_{[\,:k]}^{(b/r)} \Big(\bm{H}_{[\,:k]}^{(b/r)}\Big)^{\rm H}
  \bigg\}\right\} .
\end{align}
\hrulefill
\begin{align}\label{be2} 
 & \text{Tr}\left\{\mathcal{E}\bigg\{\Big(\bm{W}_{[k^*:\,]}^{(b/r)}\Big)^{\rm H} \bm{W}_{[k^*:\,]}^{(b/r)}
  \bm{H}_{[\,:k]}^{(b/r)}\Big(\bm{H}_{[\,:k]}^{(b/r)}\Big)^{\rm H} \bigg\}\right\} =   
  \text{Tr}\bigg\{\bigg(\nu^2 \bm{B}_{\bm{H}_{{\rm d} ~ [\,:k^*]}^{(b/r)}} + 
  \bm{\Psi}_{\widehat{\bm{H}}_{{\rm r} ~ [\,:k^*]}^{(b/r)}}\bigg)\bigg(\nu^2
  \bm{B}_{\bm{H}_{{\rm d} ~ [\,:k]}^{(b/r)}} +  \bm{\Psi}_{\bm{H}_{{\rm r} ~ [\,:k]}^{(b/r)}}\bigg)\bigg\} .
\end{align}
\hrulefill
\begin{align}\label{be3} 
 & \text{Tr}\left\{\mathcal{E}\left\{\sum\limits_{r'=2}^R \bm{V}^{(r/u)}_{r' ~ [\,:j]}
  \bigg(\sum\limits_{r'=2}^R \bm{V}^{(r/u)}_{r' ~ [\,:j]}\bigg)^{\rm H} \Big(\bm{W}_{[k:\,]}^{(r/u)}\Big)^{\rm H}
  \bm{W}_{[k:\,]}^{(r/u)}\right\}\right\} = \text{Tr}\left\{\bm{\Psi}_{\widehat{\bm{H}}^{(r/u)}_{[\,:k]}}
  \bm{\Psi}_{\Sigma\bm{V}^{(r/u)}_{r' ~ [\,:j]}}\right\} .
\end{align}
\hrulefill
\begin{align}\label{be4} 
 \bm{\Psi}_{\Sigma\bm{V}^{(r/u)}_{r' ~ [\,:j]}} =& \mathcal{E}\bigg\{\sum\limits_{r'=2}^R \bm{V}^{(r/u)}_{r' ~ [\,:j]}
  \Big(\sum\limits_{r'=2}^R \bm{V}^{(r/u)}_{r' ~ [\,:j]}\Big)^{\rm H}\bigg\} \nonumber \\
 =& \text{Bdiag}\Bigg\{\bm{0}_{M\times M},\cdots ,\bm{0}_{M\times M}, 	
  \underbrace{ \sum\limits_{r'=2}^R P_{{\rm rx},r^*/\big((r'+r^*-1)\,\text{mod}\,R\big)(j)}^{(u,d)}
  \bm{R}_{{\rm rx},r^*}^{(r/u)}}_{r^*\text{th block matrix}}, \bm{0}_{M\times M},\cdots ,
  \bm{0}_{M\times M}\Bigg\} .
\end{align}
\vspace*{-2mm}
\end{figure*}

 Fig.~\ref{FIG5} depicts the impact of the spatial correlation factor $\rho$ of RAs
 on the achievable sum-rate. Note that we assume in the simulations that both the RAs
 of the RRU and the RAs of the BBU have the same spatial correlation factor $\rho$.
 Not surprisingly, the achievable sum-rate associated with the proposed global
 optimization of power sharing is significantly higher than that without power
 sharing optimization. Most strikingly, the spatial correlation factor $\rho$ has a
 considerable negative impact on the achievable sum-rate of the system operating
 without power sharing optimization. By contrast, the spatial correlation factor
 $\rho$ hardly impacts on the system that implements the proposed global optimization
 of power sharing, except for very large $\rho \ge 0.8$. This demonstrates that the
 optimization of power sharing is capable of mitigating the detrimental impact of the
 spatial correlation.

The impact of the Rician factor $K_{Rice}$ on the achievable sum rate is
 investigated in Fig.~\ref{FIG6}. It can be seen from Fig.~\ref{FIG6} that the system relying on globally
 optimal power sharing is capable of achieving around 5.1\,bps/Hz of extra sum-rate,
 compared to the system dispensing with power sharing optimization.  Changing
 $K_{Rice}$ changes slightly the RRUs-BBU Rician channel links, which however only constitute a certain part
 of the overall system. It is therefore expected that the impact of $K_{Rice}$ on the
 achievable sum rate will not be dramatic. Indeed, it can be seen from Fig.~\ref{FIG6} that
 $K_{Rice}$ hardly has an impact on the sum rate of the system operating without power sharing optimization.
 By contrast, for the system relying on power allocation optimization, the sum rate increases
 slightly, as $K_{Rice}$ increases.

\section{Conclusions}\label{S6}

 In order to boost the fronthaul capacity of  massive MIMO aided C-RAN, we have
 proposed to globally optimize the powers shared between the pilots and data
 transmission both for the RRUs and UDs. In particular, we have derived an
 asymptotic closed-form expression of the achievable sum-rate as a function of the
 UDs' power sharing factors and the RRUs' power sharing factors. Based on this
 closed-form sum-rate expression, we have formulated the global optimal power
 sharing problem. Furthermore, by exploiting the central computing and control
 capacity of the C-RAN architecture, we have proposed to use the powerful DEA to
 solve this challenging optimization. Our extensive simulation results have
 demonstrated that the proposed global optimization of power sharing is capable
 of dramatically boosting the fronthaul capacity of massive MIMO aided C-RAN.
  Specifically,  at the configuration of 128 RAs at the BBU, the sum-rate of 10 UDs
 achieved with the optimal power sharing factors improves about 4.2\,bps/Hz to 
 5.9\,bps/Hz, when the number of RAs $M$ deployed in RRUs changes from 16 to 128.
 This represents the improvement of 33\% to 25\%, compared to the sum-rate obtained
 without optimizing power sharing factors. The influence of the key system parameters
 on the system's achievable sum-rate has also been extensively investigated. This
 study therefore has offered valuable insight and guidance concerning the practical
 deployment of massive MIMO aided C-RAN, particularly, on how to boost its fronthaul
 capacity.

\appendix

\subsection{Gallery of Lemmas}\label{Apa}

\begin{lemma}\label{L1}
 Let $X_i$ and $Y_j$ be nonnegative random variables for $1\le i\le M$ and $1\le j\le N$.
 Define $X=\sum\limits_{i=1}^M X_i$ and $Y=\sum\limits_{i=1}^N Y_i$.  Then, we have
 the following approximation of the expectation of $\log_2\left(1 + \frac{X}{Y}\right)$
 \cite{zhang2014power}:\setcounter{equation}{68}
\begin{align}\label{ae1} 
 \mathcal{E}\left\{\log_2\bigg(1 + \frac{X}{Y}\bigg)\right\} \xrightarrow[N,M\to \infty]{\text{a.s}}
  \log_2\left(1 + \frac{\mathcal{E}\{X\}}{\mathcal{E}\{Y\}}\right) .
\end{align}
\end{lemma}

\begin{lemma}[Lemma 1 in \cite{fernandes2013inter}]\label{L2}
 Let $\bm{x}\in \mathbb{C}^N$ and $\bm{y}\in \mathbb{C}^N$ be two independent random vectors,
 both having the distribution $\mathcal{CN}\big(\bm{0}_N,c \bm{I}_N\big)$.  Then,  we have
\begin{align} 
 \frac{\bm{x}^{\rm H}\bm{y}}{N} \xrightarrow[N \to \infty]{\text{a.s}}& 0 , \label{ae2} \\
 \frac{\bm{x}^{\rm H}\bm{x}}{N} \xrightarrow[N \to \infty]{\text{a.s}}& c . \label{ae3}
\end{align}
\end{lemma}

\begin{lemma}[Lemma 12 in \cite{hoydis2012random}]\label{L3}
 Let $\bm{A}\in \mathbb{C}^{N\times N}$ and $\bm{x}\sim \mathcal{CN}\big(\bm{0}_N,\frac{1}{N}\bm{I}_N\big)$.
 $\bm{A}$ has uniformly bounded spectral norm with respect to $N$ and it is independent
 of $\bm{x}$.  Then,  we have
\begin{align}\label{ae4} 
 \mathcal{E}\left\{\left|\big(\bm{x}^{\rm H}\bm{A}\bm{x}\big)^2 - \bigg(\frac{1}{N}
  \text{Tr}\big\{\bm{A}\big\}\bigg)^2\right|\right\}\xrightarrow[N \to \infty]{\text{a.s}} 0 .
\end{align}
\end{lemma}

\subsection{Derivation of $\Upsilon_{{\rm IN,3},k,j}$}\label{Apb}

 We can expand $\Upsilon_{\text{IN},3,k,j}$ as given in (\ref{be1}). Recalling Lemma~\ref{L2}
 of Appendix~\ref{Apa}, we have (\ref{be2}). Furthermore, we have (\ref{be3}). Assuming that
 the $j$th UD is served by the $r'$th RRU while it is interfering the $k$th UD served by the
 $r^*$th RRU, then $\bm{\Psi}_{\Sigma\bm{V}^{(r/u)}_{r' ~ [\,:j]}}$ is given in (\ref{be4}).
 The equations (\ref{be1}) to (\ref{be4}) are given at the bottom of the next page.

\end{document}